\documentclass[reprint, showpacs, superscriptaddress, aps, pre, twocolumn, 10pt]{revtex4-1}
\usepackage{graphicx} 
\usepackage{dcolumn}
\usepackage{color}
\usepackage{amssymb}
\usepackage{amsmath}
\usepackage{bm}	
\usepackage{float}
\usepackage{textcomp}
\usepackage{tabularx}
\usepackage[usenames,dvipsnames]{xcolor}
\usepackage{hyperref}	
\usepackage{natbib}

\hypersetup{
colorlinks,
citecolor=blue,
filecolor=blue,
linkcolor=blue,
urlcolor=blue}

\begin{document}
\title{Scaling of heat flux and energy spectrum for ``very large'' Prandtl number convection}

\author{Ambrish Pandey}
\email{pambrish@iitk.ac.in}
\affiliation{Department of Physics, Indian Institute of Technology, Kanpur 208016, India}
\author{Mahendra K. Verma}
\affiliation{Department of Physics, Indian Institute of Technology, Kanpur 208016, India}
\author{Pankaj K. Mishra}
\affiliation{Department of Chemical Physics, The Weizmann Institute of Science, Rehovot 76100, Israel}
\date{\today}

\begin{abstract}
Under the limit of infinite Prandtl number, we derive analytical expressions for the large-scale quantities, e.g., P\'{e}clet number Pe,  Nusselt number Nu, and rms value of the temperature fluctuations $\theta_\mathrm{rms}$. We complement the analytical work with direct numerical simulations, and show that $\mathrm{Nu} \sim \mathrm{Ra}^{\gamma}$ with $\gamma \approx (0.30-0.32)$,   $\mathrm{Pe} \sim \mathrm{Ra}^{\eta}$ with $\eta \approx (0.57-0.61)$, and $\theta_\mathrm{rms} \sim \mathrm{const}$. The Nusselt number is observed to be an intricate function of $\mathrm{Pe}$, $\theta_\mathrm{rms}$, and a correlation function between the vertical velocity and temperature. Using the scaling of large-scale fields, we show that the energy spectrum $E_u(k)\sim k^{-13/3}$, which is in a very good agreement with our numerical results. The  entropy spectrum $E_\theta(k)$ however exhibits dual branches consisting of $k^{-2}$ and $k^0$ spectra; the $k^{-2}$ branch corresponds to the Fourier modes $\hat{\theta}(0,0,2n)$, which are approximately $-1/(2n \pi)$.  The scaling relations for Prandtl number beyond $10^2$ match with those for infinite Prandtl number.  
\end{abstract}


\maketitle

\section{Introduction}\label{sec:intro}
Thermal convection plays a significant role in many engineering applications, as well as in natural phenomena, e.g., mantle convection, atmospheric circulation, stellar convection etc. To simplify the complex nature of convective flow, it is customary to model the flow using a simpler setup called Rayleigh-B\'{e}nard convection (RBC), in which a thin horizontal layer of the fluid is heated from below and cooled from the top~\cite{Ahlers:RMP2009, Lohse:ARFM2010}. The two non-dimensional control parameters that characterize RBC flow are the Rayleigh number Ra,  which is a measure of buoyancy force, and the  Prandtl number Pr, which is a ratio of kinematic viscosity and thermal diffusivity.  The Nusselt number Nu is defined as the ratio of the total heat flux to conductive heat flux. Experiments and numerical simulations show that the RBC flow depends quite critically on the Prandtl number. In this paper we will describe the scaling of large-scale quantities and the energy spectrum for very large and infinite Prandtl number convection. Physics of large Prandtl number convection is important for understanding convection in Earth's mantle, viscous oil, etc.

Kraichnan~\cite{Kraichnan:PF1962b} studied the scaling of Nusselt and Reynolds numbers theoretically using mixing-length theory, and deduced that $\mathrm{Nu} \sim \mathrm{Ra}^{1/3}$ for large Pr, $\mathrm{Nu} \sim (\mathrm{Pr}\mathrm{Ra})^{1/3}$ for small Pr, and $\mathrm{Nu} \sim 1$ for very small Pr.  For RBC with very large Rayleigh numbers, called the ``ultimate regime'', Kraichnan~\cite{Kraichnan:PF1962b} argued that $\mathrm{Nu} \propto (\mathrm{Ra}/ \log(\mathrm{Ra}))^{1/2}$. Grossmann and Lohse~\cite{Grossmann:JFM2000, Grossmann:PRL2001, Grossmann:PRE2002, Grossmann:PF2004, Grossmann:PF2011}, Ahlers, \textit{et al.}~\cite{Ahlers:RMP2009}, and Stevens, \textit{et al.}~\cite{Stevens:JFM2013} modelled the scaling of Nusselt and Reynolds numbers using Shraiman and Siggia's exact relations~\cite{Shraiman:PRA1990} between the dissipation rates and the Nusselt number.  For the dissipation rates, they considered the contributions from the bulk and boundary layers separately.  They showed that the parameter space (Ra, Pr) is divided into four regimes: I for $U_\mathrm{BL},\theta_\mathrm{BL}$ dominated regime; II for $U_\mathrm{bulk},\theta_\mathrm{BL}$ dominated regime; III for $U_\mathrm{BL},\theta_\mathrm{bulk}$ dominated regime; and IV for $U_\mathrm{bulk},\theta_\mathrm{bulk}$ dominated regime; here $U_\mathrm{BL}$ and $\theta_\mathrm{BL}$ are respectively the rms value of the velocity and temperature fields in the boundary layer, while  $U_\mathrm{bulk}$ and $\theta_\mathrm{bulk}$ are the corresponding quantities in the bulk.  For the infinite or very large Prandtl number (kinematic viscosity $\gg$ thermal diffusivity), most of the flow is dominated by the viscous force. Grossmann and Lohse~\cite{Grossmann:PRL2001} denote these regimes as I$_\infty$ and III$_\infty$ depending on the dominance of $\theta_\mathrm{BL}$ or $\theta_\mathrm{bulk}$ respectively.  Throughout the paper we refer to the aforementioned scaling as ``GL scaling". In this paper, we report several numerical simulations for very large and infinite Prandtl number convection; these simulations fall in either I$_\infty$ regime, or at the border of I$_\infty$ and I$_u$.  Also note that there are several mathematically rigorous bounds on the Nusselt number exponent. Whitehead and Doering~\cite{Whitehead:PRL2011} and Ierley, \textit{et al.}~\cite{Ierley:JFM2006} showed that $\mathrm{Nu} \le c \mathrm{Ra}^{5/12}$ for free-slip boundary condition, while Constantin and Doering~\cite{Constantin:JSP1999} showed that $\mathrm{Nu} \le c \mathrm{Ra}^{1/3}(1+\log(\mathrm{Ra}))^{2/3}$ in the limit of infinite Prandtl number.

A large number of experiments report the Nusselt number exponent within a range from 0.26 to 0.32 for Ra up to $10^{17}$~\cite{Castaing:JFM1989, Cioni:JFM1997, Glazier:Nature1999, Niemela:Nature2000, Ahlers:PRL2001, Niemela:JFM2003, Urban:PRL2012}. However some experiments exhibit an increase in the exponent from 0.32 to 0.37, thus signalling a signature of ultimate regime~\cite{Funfschilling:PRL2009, Ahlers:NJP2009,He:PRL2012, Chavanne:PRL1997, Roche:PRE2001}. The results from numerical simulations for small and medium Prandtl numbers and for Ra up to $10^{12}$ are consistent with the aforementioned experimental results~\cite{Kerr:JFM1996, Verzicco:JFM1999, Kerr:JFM2000, Lohse:PRL2003, Verzicco:JFM2008, Stevens:JFM2010, Stevens:JFM2011, Verma:PRE2012, Verma:IPR2013}. However, in this paper we focus on very large and infinite Prandtl number convection.

There are only a small number of experiments that investigate large Prandtl number convection. Xia, \textit{et al.}~\cite{Xia:PRL2002} studied heat transport properties of 1-pentanol, triethylene glycol,  dipropylene glycol, and water ($\mathrm{Pr}= 4-1350$), and reported the Nusselt number exponent to be in a band of 0.281 to 0.307.  Based on these observations, they claimed that $\mathrm{Nu} = 0.14\mathrm{Ra}^{0.297}\mathrm{Pr}^{-0.03}$, thus suggesting a weak dependence of the Nusselt number on the Prandtl number in the large Prandtl number regime. Lam, \textit{et al.}~\cite{Lam:PRE2002} explored the dependence of Reynolds number ($\mathrm{Re}$) on Ra and Pr and showed that $\mathrm{Re} = 0.84\mathrm{Ra}^{0.40}\mathrm{Pr}^{-0.86}$.  Since the P\'{e}clet number $\mathrm{Pe} = \mathrm{Re} \mathrm{Pr}$, Lam, \textit{et al.}~\cite{Lam:PRE2002}'s work indicates a weak dependence of Pe on Pr.

Like experiments, numerical simulations of large Prandtl number convection are limited. Silano, \textit{et al.}~\cite{Silano:JFM2010} numerically analyzed the Nusselt number for a wide range of Prandtl numbers ($\mathrm{Pr}= 10^{-1}$-$10^4$) and observed the Nusselt number exponents ranging from 2/7 for  Pr = 1 to 0.31 for $\mathrm{Pr}=10^3$. In addition, they reported that for large Pr, $\mathrm{Pe} \sim \mathrm{Ra}^{1/2}$, and a constancy of thermal fluctuations with Ra. Roberts~\cite{Roberts:GAFD1979} simulated two-dimensional RBC in the limit of $\mathrm{Pr}\rightarrow \infty$ and $\mathrm{Ra}\rightarrow \infty$ and reported that $\mathrm{Nu} \sim \mathrm{Ra}^{1/3}$ for the free-slip runs,  and $\mathrm{Nu} \sim \mathrm{Ra}^{1/5}$ for the no-slip runs. Similar results have been reported by Hansen, \textit{et al.}~\cite{Hansen:PF1990}, Schmalzl, \textit{et al.}~\cite{Schmalzl:GAFD2002}, and Breuer, \textit{et al.}~\cite{Breuer:PRE2004} in two- and three-dimensional box simulations with free-slip boundary conditions. The aforementioned researchers also studied the structures of the flow and thickness of the boundary layers.  

Another set of important quantities of interest in RBC are the energy  spectrum and entropy spectrum ($|\hat{\theta}(\mathbf k)|^2$).  For intermediate Prandtl numbers, L'vov~\cite{Lvov:PRL1991} and L'vov and Falkovich~\cite{Lvov:PD1992} predicted the Bolgiano-Obukhov scaling for small wavenumbers, but the Kolmogorov-Obukhov scaling for large wavenumbers.  We will show in this paper that the spectra for the large and infinite Prandtl number convection differ from both these laws.  Our arguments are similar to those of Batchelor~\cite{Batchelor:JFM1959a} for small Prandtl number, except that the role of temperature and velocity is reversed in our work since the Prandtl number is large.

As discussed above, there are significant development in the understanding of large and infinite Prandtl number convection, especially from the GL scaling~\cite{Grossmann:PRL2001}.  Still, many issues remain unresolved in this field.  We address some of these issues using analytical approach and direct numerical simulation.   Under the limit of infinite Prandtl number, we derive a linear relationship between the velocity and temperature fields, which enables us to derive interesting exact relations for the P\'{e}clet  and Nusselt numbers.  We also derive energy and entropy spectra analytically using the above relations, as well as the temperature equation.  In addition, we also obtain analytic expressions for the thermal fluctuations, and viscous and thermal dissipation rates.  The Nusselt number and thermal dissipation rates are intricately dependent on the Ra-dependent correlation between the vertical velocity and the temperature fluctuations.  These relations provide valuable insights into large-Pr convective turbulence. It is important to note that our theoretical work are based on the dimensional and scaling analysis of the equations for the velocity and temperature fields.  Our approach differs somewhat from the GL scaling, which is based on Shraiman and Siggia's~\cite{Shraiman:PRA1990} exact relation, and the scaling of the dissipation rates in the bulk and boundary layers.

To validate our analytical predictions, we perform direct numerical simulations of RBC flows for $\mathrm{Pr}=10^2,10^3, \infty$ and Rayleigh numbers varying from $10^4$ to $10^8$.  Using the numerical data we compute scaling for the large-scale quantities like the Nusselt and P\'{e}clet numbers,  the rms fluctuations of the temperature field, the viscous and thermal dissipation rates, and compare them with analytic predictions.  We also compute energy and entropy spectra using the numerical data.  We find that our numerical results are in good agreement with the analytical predictions, as well as with the GL scaling.

The structure of paper is as follows: In \textsection~\ref{sec:eqns}, we present the governing equations and analytical  expressions for various large-scale quantities like the Nusselt and P\'{e}clet numbers, and the viscous and thermal dissipation rates.  In \textsection~\ref{sec:numerical} we  discuss the details of our numerical simulations. Scaling relations of Pe, Nu, and dissipation rates are  discussed in \textsection~\ref{sec:results}. In \textsection~\ref{sec:spectrum} we derive energy and entropy spectra using the dynamical equations in the  $\mathrm{Pr}= \infty$ limit, and  complement them with the numerical results for infinite and large Prandtl number simulations.   We conclude in \textsection~\ref{sec:conclusion}.

\section{Governing equations and analytic computations} \label{sec:eqns}
The dynamical equations of RBC under Boussinesq approximations are
\begin{eqnarray}
\frac{\partial \bf u}{\partial t} + (\bf u \cdot \nabla) \bf u & = & -\frac{\nabla p}{\rho_0} + \alpha g \theta \hat{z} + \nu \nabla^2 \bf u , \label{eq:u_dim} \\
\frac{\partial \theta}{\partial t} + (\bf u \cdot \nabla) \theta & = & \frac{\Delta}{d} u_z + \kappa \nabla^2 \theta , \label{eq:th_dim} \\
\nabla \cdot \bf u & = & 0,
\end{eqnarray}
where $\mathbf u = (u_x,u_y,u_z)$ is the velocity field, $p$ and $\theta$ are respectively the deviations of pressure and temperature from the heat conduction state, $\rho_0$ is the mean density of fluid, $\alpha, \nu$, and $\kappa$ are respectively the thermal expansion coefficient, kinematic viscosity, and thermal diffusivity of fluid, $g$ is the acceleration due to gravity, $\hat{z}$ is the buoyancy direction, and $\Delta$ is the temperature difference between the two plates kept apart by a vertical distance $d$. 

For large and infinite Prandtl number convection, it is customary to nondimensionalize the above set of equations using $\sqrt{\alpha g \Delta d/\mathrm{Pr}}$, $d$, and $\Delta$ as the velocity, length, and temperature scales respectively~\cite{Silano:JFM2010}, which yields
\begin{eqnarray}
\frac{1}{\mathrm{Pr}}\left[ \frac{\partial \bf u}{\partial t} + (\bf u \cdot \nabla) \bf u \right] & = & -\nabla \sigma + \theta \hat{z} + \frac{1}{\sqrt{\mathrm{Ra}}} \nabla^2 \bf u, \label{eq:u_largeP} \\
\frac{\partial \theta}{\partial t} + (\bf u \cdot \nabla) \theta & = & u_z + \frac{1}{\sqrt{\mathrm{Ra}}} \nabla^2 \theta , \label{eq:th_largeP} \\
\nabla \cdot \bf u & = & 0, \label{eq:cont}
\end{eqnarray}
where $\mathrm{Ra}= \alpha g \Delta d^3 /\nu \kappa$, $\mathrm{Pr}= \nu/\kappa$, and $\sigma=p/\mathrm{Pr}$~\cite{Silano:JFM2010, Schmalzl:GAFD2002}. Under the limit of $\mathrm{Pr}=\infty$, the momentum equation gets simplified to~\cite{Breuer:EPL2009}
\begin{equation}
-\nabla \sigma + \theta \hat{z} + \frac{1}{\sqrt{\mathrm{Ra}}} \nabla^2 {\bf u} = 0. \label{eq:u_infty_nondim}
\end{equation}
The equations for the temperature field and the incompressibility condition remain unchanged.  Note that the pressure term plays an important role in infinite Prandtl number convection, and it cannot be ignored.  An easy inspection reveals that the $\nabla \cdot {\bf u} = 0$ condition would be violated in the absence of the pressure term.

The momentum equation in a dimensional form is
\begin{equation}
-\nabla \sigma + \alpha g \theta \hat{z} + \nu \nabla^2 {\bf u} = 0. \label{eq:u_infty}
\end{equation}
The above equation is linear, hence analytically tractable.  It is more convenient to analyze the above equation in Fourier space, which is
\begin{eqnarray}
-i {\mathbf k} \hat{\sigma}(\mathbf k) + \alpha g \hat{\theta}(\mathbf k) \hat{z}- \nu k^2 \hat{\mathbf u}(\mathbf k) & = & 0, \label{eq:u_k}
\end{eqnarray}
where ${\mathbf k}=(k_x,k_y,k_z)$ is the wavenumber, and $\hat{f}(\mathbf k)$ is the Fourier mode of the field $f$.  Using the incompressible condition 
(${\mathbf k} \cdot \hat{\mathbf u}(\mathbf k) = 0$), we deduce that 
\begin{eqnarray}
\hat{\sigma}(\mathbf k) & = & -i \frac{k_z}{k^2} \alpha g \hat{\theta}(\mathbf k), \label{eq:p_k} \\
\hat{u}_z (\mathbf k) & = & \frac{\alpha g}{\nu}  \frac{k_\perp^2}{k^4} \hat{\theta}(\mathbf k), \label{eq:u_z} \\
\hat{u}_{x,y} (\mathbf k) & = & - \frac{\alpha g}{\nu}  \frac{k_z k_{x,y}}{k^4} \hat{\theta}(\mathbf k), \label{eq:u_xy}
\end{eqnarray}
where $k_\perp^2 = k_x^2 + k_y^2$. The Fourier modes in general do not satisfy the boundary conditions at the plates.  Yet, they capture the large-scale modes quite accurately since the energy of the convective flow is dominated by these modes. This feature becomes more significant for infinite Prandtl number since  the amplitudes of the larger Fourier modes decrease steeply with the wavenumber ($\sim k^{-2}$), as evident from the aforementioned equations.

Using the above equations, we can derive the total kinetic energy as
\begin{equation}
E_u = \frac{1}{2} u^2 = \frac{1}{2} \sum_\mathbf k |\hat{u}(\mathbf k)|^2 = \frac{1}{2}\left(\frac{\alpha g}{\nu}\right)^2 \sum_\mathbf k|\hat{\theta}(\mathbf k)|^2 \frac{k_\perp^2}{k^6}.
\end{equation} 
The total energy is dominated by the large-scale flows.   Therefore, the P\'{e}clet number Pe is
\begin{equation}
\mathrm{Pe} = \frac{U_L d}{\kappa} = \frac{d}{\kappa} \sqrt{2 E_u} = \frac{\alpha g d}{\nu \kappa} \left(\sum_\mathbf k|\hat{\theta}(\mathbf k)|^2 \frac{k_\perp^2}{k^6} \right)^{1/2}. \label{eq:Eu}
\end{equation} 
In terms of nondimensional parameters
\begin{equation}
\mathrm{Pe} =  \mathrm{Ra} \left( \sum_\mathbf k|\hat{\theta}(\mathbf k)|^2 \frac{k_\perp^2}{k^6} \frac{1}{ d^4 \Delta^2} \right)^{1/2}. \label{eq:Pe}
\end{equation} 

One of the generic features of thermal convection in a box is the finite amplitude of $\hat{\theta}(0,0,2n)$ Fourier modes for small $n$, e.g., $n=1,2,3$. Mishra and Verma~\cite{Mishra:PRE2010} showed using arguments based on energy transfers that
\begin{equation}
\hat{\theta}(0,0,2n) \approx -\frac{\Delta}{2 n \pi}.
\end{equation} 
The $\hat{\theta}(0,0,2n)$ modes  play an important role in determining the vertical profile of temperature. The temperature averaged over horizontal planes drops sharply near the plates (in the boundary layer), and it is approximately a constant in the bulk.  We will show later in the paper that the temperature drop near the plates gets significant contributions from the $\hat{\theta}(0,0,2n)$ modes~\cite{Verma:IPR2013}.  

We will demonstrate later in the paper that the $\hat{\theta}(0,0,2n)$  modes dominate the temperature fluctuations for large and infinite Prandtl number convection. Thus, 
\begin{equation}
\theta_L \approx \theta_{\mathrm{rms}} \approx \sqrt{2 E_\theta} \approx \Delta ,
\end{equation} 
where $E_\theta$ is defined as
\begin{equation}
E_\theta = \frac{1}{2} \theta^2 = \frac{1}{2} \sum_\mathbf k |\hat{\theta}(\mathbf k)|^2.
\end{equation}
To quantify the contributions from other thermal modes, we define a residual temperature fluctuation $\theta_\mathrm{res}$ as
\begin{equation}
\theta_{\mathrm{res}}^2 = \theta^2 - \sum_n |\hat{\theta}(0,0,2n)|^2. \label{eq:res_th}
\end{equation}
It is important to note that $\hat{u}_z(0,0,n)=0$ due to the absence of net mass flux across any horizontal cross-section in the box.  As a result, the  $\hat{\theta}(0,0,2n)$ modes do not contribute to the heat flux $H$, which is 
\begin{equation}
H \propto \langle u_z \theta \rangle = \sum_{\mathbf k} (\hat{u}_z(\mathbf k) \hat{\theta}^*(\mathbf k) + \hat{u}^*_z(\mathbf k) \hat{\theta}(\mathbf k)),
\end{equation}
where, $u_z^*$ and $\theta^*$ denote respectively the complex conjugate of the vertical velocity field and the temperature field. Thus, the heat flux gets contributions only from $\theta_{\mathrm{res}}$ fluctuations.  

The viscous and thermal dissipation rates  provide important information about the scaling of large-scale quantities~\cite{Shraiman:PRA1990, Grossmann:JFM2000, Emran:JFM2008, Emran:EPJE2012}. Shraiman and Siggia~\cite{Shraiman:PRA1990} relate these dissipation rates to the Nusselt number using the following exact relationships:
\begin{eqnarray}
\epsilon_u & = & \nu \langle |\nabla \times \mathbf u|^2 \rangle = \frac{\nu^3}{d^4} \frac{(\mathrm{Nu}-1)\mathrm{Ra}}{\mathrm{Pr}^2}, \label{eq:eps_u} \\
\epsilon_T & = & \kappa \langle |\nabla T|^2 \rangle =  \kappa \frac{\Delta^2}{d^2} \mathrm{Nu}, \label{eq:eps_th}  
\end{eqnarray}
where $T$ is the temperature field defined as $T = T_{\mathrm{conduction}} + \theta$. In the viscous regime, we use  $\epsilon_u \propto \nu U_L^2/d^2$ and define a  normalized viscous dissipation rate as
\begin{equation}
C_{\epsilon_u} = \frac{\epsilon_u}{\nu U_L^2/d^2} = \frac{(\mathrm{Nu}-1)\mathrm{Ra}}{\mathrm{Pe}^2}. \label{eq:C_eps_u}
\end{equation}
We define a normalized thermal dissipation rate as
\begin{equation}
C_{\epsilon_T,1} = \frac{\epsilon_T}{\kappa \Delta^2/d^2} = \mathrm{Nu}. \label{eq:C_eps_th_1}
\end{equation}
Note that  $\epsilon_u \propto U_L^3/d$ in turbulent regime, but this formula is not applicable for the large and infinite Prandtl number convection because of the laminar nature of the flow.  The temperature equation however is fully nonlinear, hence $\epsilon_T \propto U_L \theta_L^2/d$~\cite{Grossmann:PRL2001, Emran:JFM2008, Emran:EPJE2012, Verma:IPR2013}. Therefore we define another normalized thermal dissipation rate as 
\begin{equation}
C_{\epsilon_T,2} = \frac{\epsilon_T}{U_L \theta_L^2/d} = \frac{\mathrm{Nu}}{\mathrm{Pe}} \left( \frac{\Delta}{\theta_L} \right)^2. \label{eq:C_eps_th_2}
\end{equation}
We will compute these dissipation rates using numerical data and use them for validation, as well as for the prediction of Nusselt number scaling. 

\section{Numerical Method} \label{sec:numerical}
We solve the nondimensionalized RBC equations [Eqs.~(\ref{eq:u_largeP}, \ref{eq:th_largeP}, \ref{eq:cont})] numerically for both the free-slip and no-slip boundary conditions.  We simulate free-slip RBC flow in a three-dimensional box of dimension $2\sqrt{2}:2\sqrt{2}:1$ using a pseudospectral code Tarang~\cite{Verma:Pramana2013}. On the top and bottom plates we employ free-slip and isothermal conditions for the velocity and temperature fields respectively. However, periodic boundary condition is employed on the lateral walls.   Fourth-order Runge-Kutta (RK4) method is used for  time advancement, and the 2/3 rule for dealiasing.  

To complement the aforementioned free-slip numerical runs, we also simulate RBC flow under no-slip boundary condition in a two-dimensional box of aspect ratio one.  We used the spectral element code NEK5000~\cite{Fischer:JCP1997} for this purpose. Runs were performed in a box with $28 \times 28$ spectral elements with 7th-order polynomials within each element, resulting in a $196^2$ effective grid points in the box.  For the spectrum study however we used 15th-order polynomial, which yields  $420^2$ effective grid points in the box.  We will describe below that the two-dimensional no-slip and three-dimensional free-slip runs exhibit similar results for the large-scale quantities and energy spectrum.   

\begin{figure}
\begin{center} 
\includegraphics[scale=0.25]{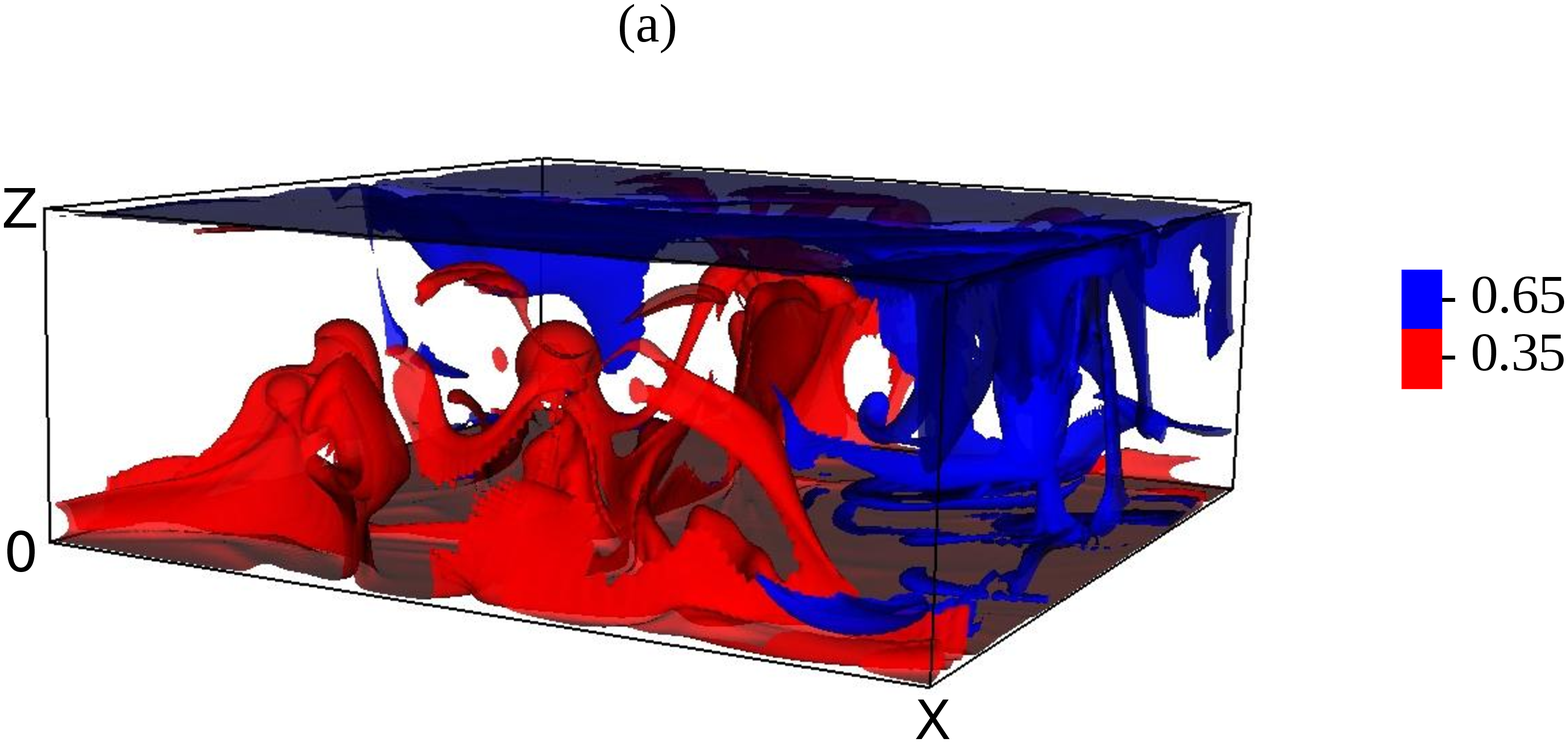}
\includegraphics[scale=0.25]{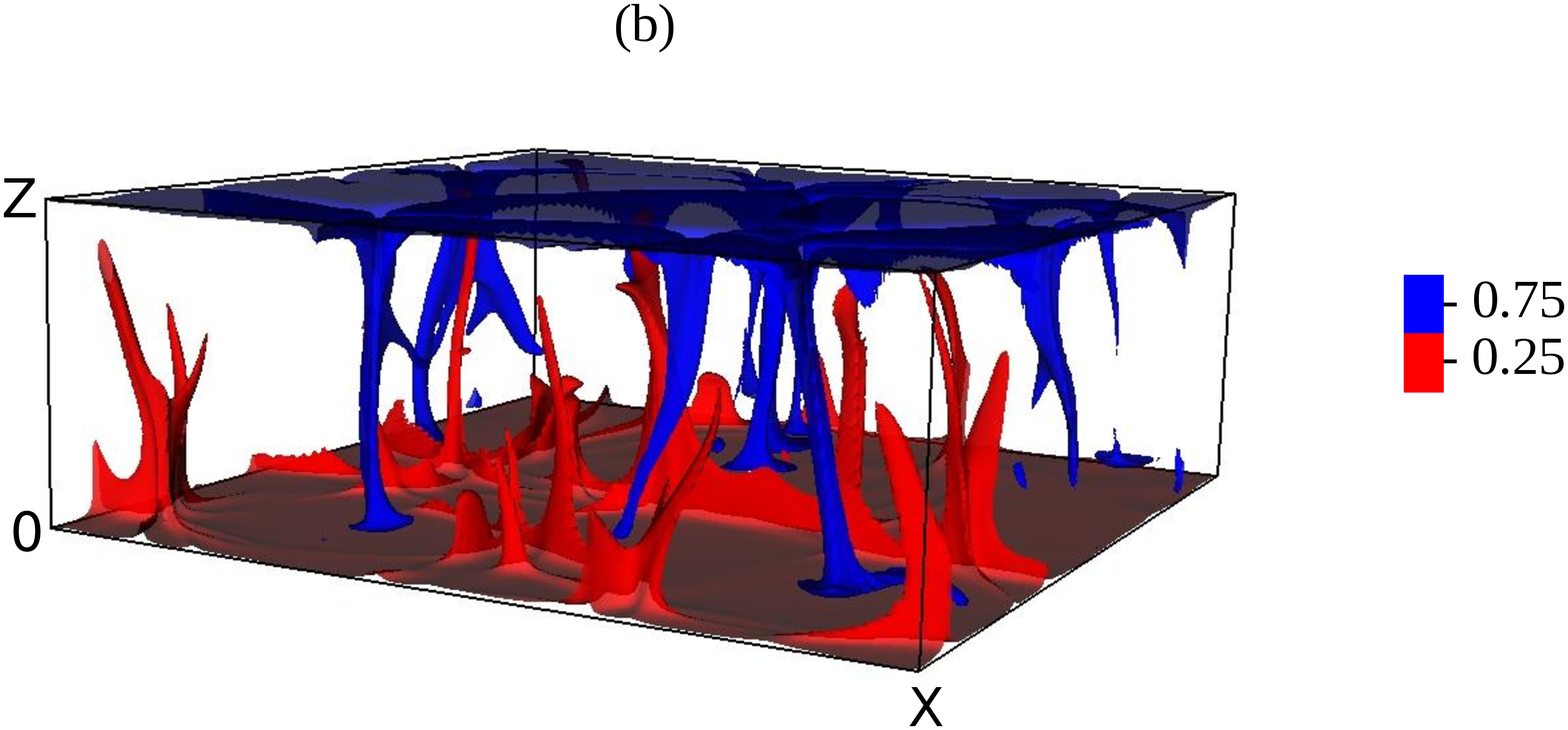}
\includegraphics[scale=0.25]{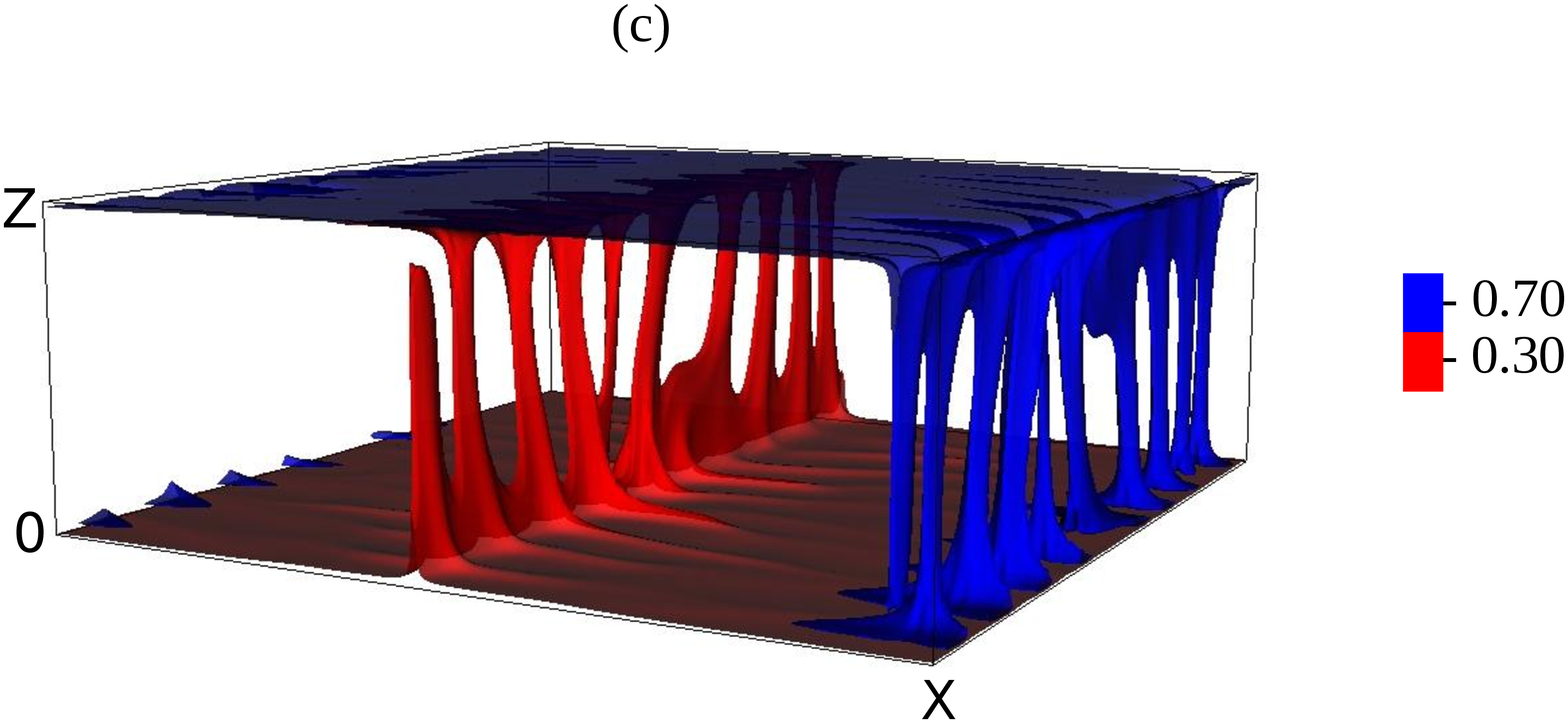}
\end{center}
\caption{(Color online) Temperature isosurfaces for (a) $\mathrm{Pr}= 10^2$ and $\mathrm{Ra}= 10^7$, (b) $\mathrm{Pr}= 10^3$ and $\mathrm{Ra}= 6 \times 10^6$, and (c) $\mathrm{Pr} = \infty$ and $\mathrm{Ra} = 6.6 \times 10^6$. The red structures represent hot plumes, and blue ones represent cold plumes. The large-scale structures get sharper with the increase of Pr.} 
\label{fig:temp_contour}
\end{figure}

We perform simulations for $\mathrm{Pr}= 10^2$, $10^3$, $\infty$, and  $\mathrm{Ra}$ in the range from $6 \times 10^4$ to  $1 \times 10^8$.   For such large Prandtl numbers, the  kinematic viscosity is much larger than the thermal diffusion coefficient, consequently coherent thin plumes are generated in such flows~\cite{Schmalzl:GAFD2002, Hansen:PF1990,Breuer:PRE2004,Stevens:NJP2010}. Fig.~\ref{fig:temp_contour} illustrates the temperature isosurfaces  of the flow structures for ($\mathrm{Pr}= 10^2, \mathrm{Ra}= 10^7$),  ($\mathrm{Pr}= 10^3, \mathrm{Ra}= 6\times 10^6$), and ($\mathrm{Pr}= \infty, \mathrm{Ra}= 6.6 \times 10^6$). The figures demonstrate that the plumes become thinner and sharper with the increase of Prandtl number. Earth's mantle that has very large Prandtl number ($\mathrm{Pr} \approx 10^{25}$) shows similar structures. A cross-sectional view of the flow pattern exhibits spoke like patterns, first shown by Busse and Whitehead~\cite{Busse:JFM1974} in their experiments with silicone oil. Also, $k_{\mathrm{max}} \eta_u \gtrsim 1$  and $k_{\mathrm{max}} \eta_\theta \gtrsim 1$ for all of our simulation runs; here $\eta_u = (\nu^3/\epsilon_u)^{1/4}$ and $\eta_\theta = (\kappa^3/\epsilon_u)^{1/4}$ are the Kolmogorov length and Batchelor length for the velocity and temperature fields respectively. Thus our simulations are numerically well resolved.

\begin{table*}
\begin{center}
\caption{Details of numerical simulations performed for free-slip boundary condition. The table contains P\'{e}clet number Pe, Nusselt number Nu; the normalized viscous dissipation rate: numerically computed $C_{\epsilon_u}^{\mathrm{comp.}} = \epsilon_u/(\nu U^2_L /d^2)$, and estimated $C_{\epsilon_u}^{\mathrm{est.}} = (\mathrm{Nu}-1)\mathrm{Ra}/\mathrm{Pe}^2$; the normalized thermal dissipation rates: numerically computed $C_{\epsilon_{_T},1}^{\mathrm{comp.}}=  \epsilon_{_T}/(\kappa \Delta^2 /d^2)$, $C_{\epsilon_{_T},2}^{\mathrm{comp.}} = \epsilon_{_T}/(U_L \theta^2_L /d)$, and estimated $C_{\epsilon_{_T},2}^{\mathrm{est.}} = (\mathrm{Nu}/\mathrm{Pe})(\Delta/\theta_L)^2$. It also lists globally averaged $k_\mathrm{max} \eta_u$  and $k_\mathrm{max} \eta_\theta$, which are $\gtrapprox 1$ for all the cases.}
\begin{tabular*}{\textwidth}{p{1.2cm} p{2cm} p{1.2cm} p{1.2cm} p{2cm} p{1.2cm} p{1.2cm} p{1.2cm} p{1.2cm} p{1.2cm} p{1.2cm} p{1.2cm}} 
\hline \hline \\[1 pt]
 Pr & Ra & Grid & Nu & Pe & $C_{\epsilon_u}^{\mathrm{comp.}}$ & $C_{\epsilon_u}^{\mathrm{est.}}$ & $C_{\epsilon_{_T},1}^{\mathrm{comp.}}$ & $C_{\epsilon_{_T},2}^{\mathrm{comp.}}$ & $C_{\epsilon_{_T},2}^{\mathrm{est.}}$ & $k_{\mathrm{max}}\eta_u$ & $k_{\mathrm{max}}\eta_{\theta}$ \\[2 mm]
\hline\\[1 pt]
	$10^2$ & $1.0 \times 10^5$ &	 $256^3$ & 9.8 & $1.98 \times 10^2$ & 22.3 & 22.3 & 9.8 & 0.60 & 0.60 & 61.4 & 1.9 \\
	$10^2$ & $2.0 \times 10^5$ &  $256^3$ & 11.2 & $2.92 \times 10^2$ & 24.0 & 23.9 & 11.2 & 0.47 & 0.46 & 50.0 & 1.6 \\
	$10^2$ & $6.5 \times 10^5$ & $256^3$ & 17.3 & $6.15 \times 10^2$ & 28.6 & 28.3 & 17.5 & 0.36 & 0.34 & 32.0 & 1.0 \\
	$10^2$ & $2.0 \times 10^6$ & $512^3$ & 24.1 & $1.20 \times 10^3$ & 32.1 & 32.2 & 24.1 & 0.25 & 0.24 & 45.7 & 1.4 \\
	$10^2$ & $5.0 \times 10^6$ & $512^3$ & 31.0 & $1.96 \times 10^3$ & 39.5 & 39.1 & 30.9 & 0.19 & 0.19 & 33.9 & 1.1 \\
	$10^2$ & $1.0 \times 10^7$ & $1024^3$ & 38.1 & $2.92 \times 10^3$ & 43.7 & 43.4 & 38.2 & 0.16 & 0.16 & 54.3 & 1.7 \\
		
	$10^3$ & $6.5 \times 10^4$ & $256^3$ & 8.6 & $1.53 \times 10^2$ & 21.4 & 21.4 & 8.6 & 0.69 & 0.68 & 223 & 1.3 \\
	$10^3$ & $1.0 \times 10^5$ & $256^3$ & 9.8 & $1.98 \times 10^2$ & 22.3 & 22.3 & 9.8 & 0.60 & 0.60 & 194 & 1.1 \\
	$10^3$ & $2.0 \times 10^5$ & $512^3$ & 12.1 & $3.01 \times 10^2$ & 23.9 & 23.9 & 12.1 & 0.48 & 0.48 & 309 & 1.7 \\
	$10^3$ & $3.2 \times 10^5$ & $512^3$ & 14.1 & $3.98 \times 10^2$ & 27.2 & 27.1 & 14.1 & 0.42 & 0.43 & 261 & 1.5\\
	$10^3$ & $2.0 \times 10^6$ & $1024^3$ & 24.3 & $1.10 \times 10^3$ & 38.7 & 38.3 & 24.3 & 0.26 & 0.26 & 277 & 1.6\\
	$10^3$ & $6.0 \times 10^6$ & $1024^3$ & 34.2 & $2.13 \times 10^3$ & 43.4 & 43.7 & 34.2 & 0.19 & 0.19 & 200 & 1.1\\

	$\infty$ & $7.0 \times 10^4$ & $128^3$ & 8.8 & $1.59 \times 10^2$ & 21.4 & 21.6 & 8.8 & 0.67 & 0.68 & $\infty$ & 1.7 \\
	$\infty$ & $1.9 \times 10^5$ & $128^3$ & 12.1 & $3.02 \times 10^2$ & 23.9 & 24.0 & 12.0 & 0.48 & 0.49 & $\infty$ & 2.4 \\
	$\infty$ & $3.2 \times 10^5$ & $128^3$ & 14.1 & $4.14 \times 10^2$ & 25.1 & 25.1 & 14.1 & 0.41 & 0.42 & $\infty$ & 2.0 \\
	$\infty$ & $6.5 \times 10^5$ & $128^3$ & 17.4 & $6.36 \times 10^2$ & 26.7 & 26.7 & 17.4 & 0.33 & 0.34 & $\infty$ & 1.6 \\
	$\infty$ & $3.9 \times 10^6$ & $256^3$ & 30.3 & $1.95 \times 10^3$ & 30.3 & 30.4 & 30.3 & 0.19 & 0.19 & $\infty$ & 1.8 \\
	$\infty$ & $6.5 \times 10^6$ & $256^3$ & 36.1 & $2.70 \times 10^3$ & 33.5 & 31.8 & 36.0 & 0.16 & 0.16 & $\infty$ & 1.5 \\
	$\infty$ & $9.8 \times 10^6$ & $256^3$ & 41.2 & $3.34 \times 10^3$ & 35.8 & 35.6 &  41.1 & 0.15 & 0.15 & $\infty$ & 1.3 \\
	$\infty$ & $1.9 \times 10^7$ & $256^3$ & 51.2 & $5.20 \times 10^3$ & 36.8 & 36.6 & 51.2 & 0.12 & 0.12 & $\infty$ & 1.1 \\
	$\infty$ & $1.0 \times 10^8$ & $512^3$ & 87.5 & $1.38 \times 10^4$ & 45.6 & 45.3 & 87.2 & 0.07 & 0.07 & $\infty$ & 1.3 \\

\hline \hline
\end{tabular*}
\label{table:pr_inf}
\end{center} 
\end{table*}

\begin{table}
\begin{center}
\caption{Details of RBC simulations with no-slip boundary condition for a two-dimensional box of aspect ratio one. }
\begin{tabular}{p{1.5cm} p{2cm} p{1.5cm} p{1.5cm} p{1.5cm}}
\hline \hline \\[1 pt]
Pr & Ra & Grid & Nu & Pe \\[2 mm]
\hline \\[1 pt]
$10^2$ & $1 \times 10^4$ & $196^2$ & 2.2 & $1.43 \times 10^1$ \\
$10^2$ & $1 \times 10^5$ & $196^2$ & 3.9 & $5.73 \times 10^1$ \\
$10^2$ & $1 \times 10^6$ & $196^2$ & 7.1 & $1.93 \times 10^2$ \\
$10^2$ & $1 \times 10^7$ & $420^2$ & 14.4 & $7.55 \times 10^2$ \\
$10^2$ & $5 \times 10^7$ & $196^2$ & 22.6 & $1.99 \times 10^3$ \\
\hline \hline
\end{tabular}
\label{table:no_slip}
\end{center}
\end{table}

We compute various global quantities (e.g., $\theta_L$, P\'{e}clet and Nusselt numbers), and energy and entropy spectra using the numerical data generated by our simulations.  These quantities are averaged over 200-300 eddy turnover time after the flow has reached a steady state.  Note that the system takes around a thermal diffusive time to reach a steady state. For better statistical averaging, the Nusselt number is computed by averaging the heat flux over the box volume. For all our runs,  the number of grid points in the thermal boundary layers are greater than 5 to 6,  which is  consistent with the Gr\"{o}tzbach criteria~\cite{Grotzbach:JCP1983}. For example, for Pr = $\infty$ and Ra = $10^8$, the thermal boundary layers at both the plates contain 10 points. 

Table~\ref{table:pr_inf} exhibits details of our free-slip numerical simulations. In the table, we list the normalized dissipation rates computed using the numerical data and compare them with the ones derived using the exact relations (using the Nusselt number). The estimated values are in very good agreement with the numerically computed ones, thus validating our numerical simulations. We however remark that the viscous and thermal dissipation rates exhibit temporal and spatial variability, as shown by many researchers.  For example, Emran and Schumacher~\cite{Emran:JFM2008, Emran:EPJE2012} performed a detailed  numerical analysis of the thermal dissipation rate $\epsilon_T$  and showed that the scaling of $\epsilon_T$  in boundary layer and in bulk are different.

The details of our no-slip runs are exhibited in Table~\ref{table:no_slip}. We also ensured grid-independence of our numerical 
program by performing simulations on grids with higher and lower resolutions, and comparing our results.  The global quantities like P\'{e}clet and Nusselt numbers were found to be within 1-2\% for these simulation.

In the next section we will describe the  scaling of large-scale quantities derived using the numerical data.

\section{Scaling of large-scale quantities} \label{sec:results}
In this section we report average values of the P\'{e}clet and Nusselt numbers, as well as that of the temperature fluctuations and dissipation rates, for various Prandtl and Rayleigh numbers.  These values are compared with the analytical predictions.

\subsection{Temperature fluctuations} \label{subsec:temp_fluct}
\begin{figure}
\begin{center}
\includegraphics[scale=0.36]{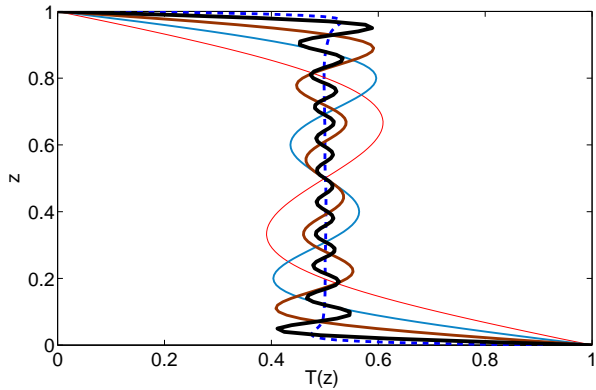}
\end{center}
\caption{(Color online) Averaged temperature $T(z)$ (dashed blue curve)  as a function of $z$ for $\mathrm{Pr}= \infty$  and $\mathrm{Ra}= 10^8$. The mean temperature remains almost a constant in the bulk, and it displays a sharp gradient near the top and bottom plates. The red, sky blue, brown, and black curves (curves with increasing thickness) represent $T_c+\sum_n 2 \hat{\theta}(0,0,2n) \sin(2\pi n z)$ for $n=1,2,4$, and 10 respectively; here $T_c=1-z$ is the conduction profile. The curve for $n=10$ matches quite well with $T(z)$.}  
\label{fig:temp_inf}
\end{figure}

For large Prandtl number convection, Silano, \textit{et al.}~\cite{Silano:JFM2010} showed that the temperature fluctuations  are independent of Ra and Pr. Here we analyze this issue in more detail. As discussed in the previous section, $\hat{\theta}(0,0,2n)$ modes play an important role in turbulent convection.  We compute $\hat{\theta}(0,0,2n)$ modes for small $n$ using the numerical data of our simulation. These values, exhibited in Table~\ref{table:th_2n} for some typical parameters, are  in good agreement with the predictions of Mishra and Verma~\cite{Mishra:PRE2010} that $\hat{\theta}(0,0,2n) \approx -\Delta/(2 n \pi)$.  In Fig.~\ref{fig:temp_inf} we plot the averaged temperature profile $\bar{T}(z)$,  as well as $1-z+\sum_n 2 \hat{\theta}(0,0,2n) \sin(2\pi nz)$ for $n=1,2,4$ and 10 (note that $T_\mathrm{conduction} = 1-z$). The figure demonstrates that the $\bar{T}(z)$ is well approximated by $1-z+\sum_{n=0}^{10} 2 \hat{\theta}(0,0,2n) \sin(2\pi nz)$. Hence we conclude that the $\hat{\theta}(0,0,2n)$ modes 
contribute significantly to $\bar{T}(z)$.

\begin{table}
\begin{center}
\caption{Numerically computed values of the Fourier modes $\hat{\theta}(0,0,2)$, $\hat{\theta}(0,0,4)$, $\hat{\theta}(0,0,6)$, and  $\hat{\theta}(0,0,8)$ for $\mathrm{Pr}= 10^2, 10^3,$ and $\infty$. The values obtained from our simulation are in good agreement with Mishra and Verma's~\cite{Mishra:PRE2010} theoretical prediction  that $\hat{\theta}(0,0,2n)$ $\approx -1/2\pi n$ (the last row of the table).}
\begin{tabular}{p{1cm} p{1.5cm} p{1.4cm} p{1.4cm} p{1.4cm} p{1cm}}
\hline \hline \\[1 pt]
Pr & Ra & $\hat{\theta}(0,0,2)$ & $\hat{\theta}(0,0,4)$ & $\hat{\theta}(0,0,6)$ & $\hat{\theta}(0,0,8)$ \\[2 mm]
\hline \\[1 pt]
$10^2$ & $1 \times 10^7$ & -0.16 & -0.081 & -0.054 & -0.040 \\
$10^3$ & $6 \times 10^6$ & -0.16 & -0.082 & -0.055 & -0.040 \\
$\infty$ & $1 \times 10^8$ & -0.16 & -0.080 & -0.054 & -0.041 \\ 
- & $-1/2\pi n$ & -0.16 & -0.080 & -0.053 & -0.039 \\
\hline \hline
\end{tabular}
\label{table:th_2n}
\end{center}
\end{table}

\begin{figure}
\begin{center}
\includegraphics[scale = 0.23]{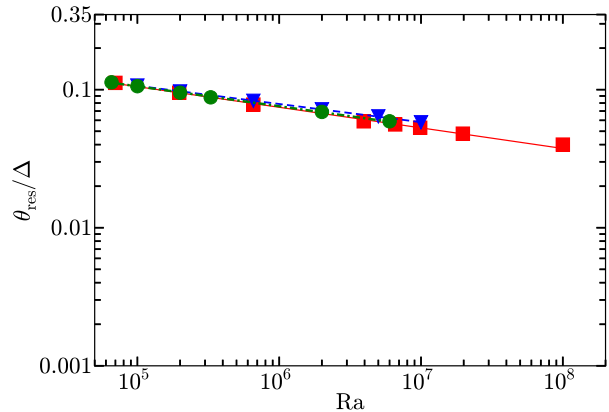}
\end{center}
\caption{(Color online) Normalized residual temperature fluctuation ($\theta_{\mathrm{res}}/\Delta$) as a function of Ra for free-slip runs. The data for $\mathrm{Pr}= \infty$ (red squares), $\mathrm{Pr}= 10^2$ (blue down-pointing triangles), and $\mathrm{Pr}= 10^3$ (green  circles) collapse to a single function $\theta_\mathrm{res} \approx a_1 \Delta \mathrm{Ra}^{-\delta}$.  The prefactors and exponents for the three runs are approximately equal.}
\label{fig:th_ra_2}
\end{figure}

We also compute the residual temperature fluctuations $\theta_\mathrm{res}$ defined using Eq.(\ref{eq:res_th}), and  observe that 
\begin{equation}
\theta_\mathrm{res} =  a_1 \Delta \mathrm{Ra}^{-\delta}, \label{eq:theta_res_delta}
\end{equation}
as shown in Fig.~\ref{fig:th_ra_2}.  We deduce that $\theta_{\mathrm{res}}/\Delta = (0.59 \pm 0.08)\mathrm{Ra}^{-0.15 \pm 0.01}$ for $\mathrm{Pr}= \infty$,   $\theta_{\mathrm{res}}/\Delta = (0.49 \pm 0.02)\mathrm{Ra}^{-0.13 \pm 0.01}$ for $\mathrm{Pr}=10^2$, and $\theta_{\mathrm{res}}/\Delta = (0.56 \pm 0.04)\mathrm{Ra}^{-0.14 \pm 0.01}$ for $\mathrm{Pr}=10^3$. Thus, the scaling exponents as well as the prefactors of $\theta_\mathrm{res}$ for various Prandtl numbers are nearly same for $\mathrm{Pr}\ge 10^2$. 

\begin{figure}
\begin{center}
\includegraphics[scale = 0.23]{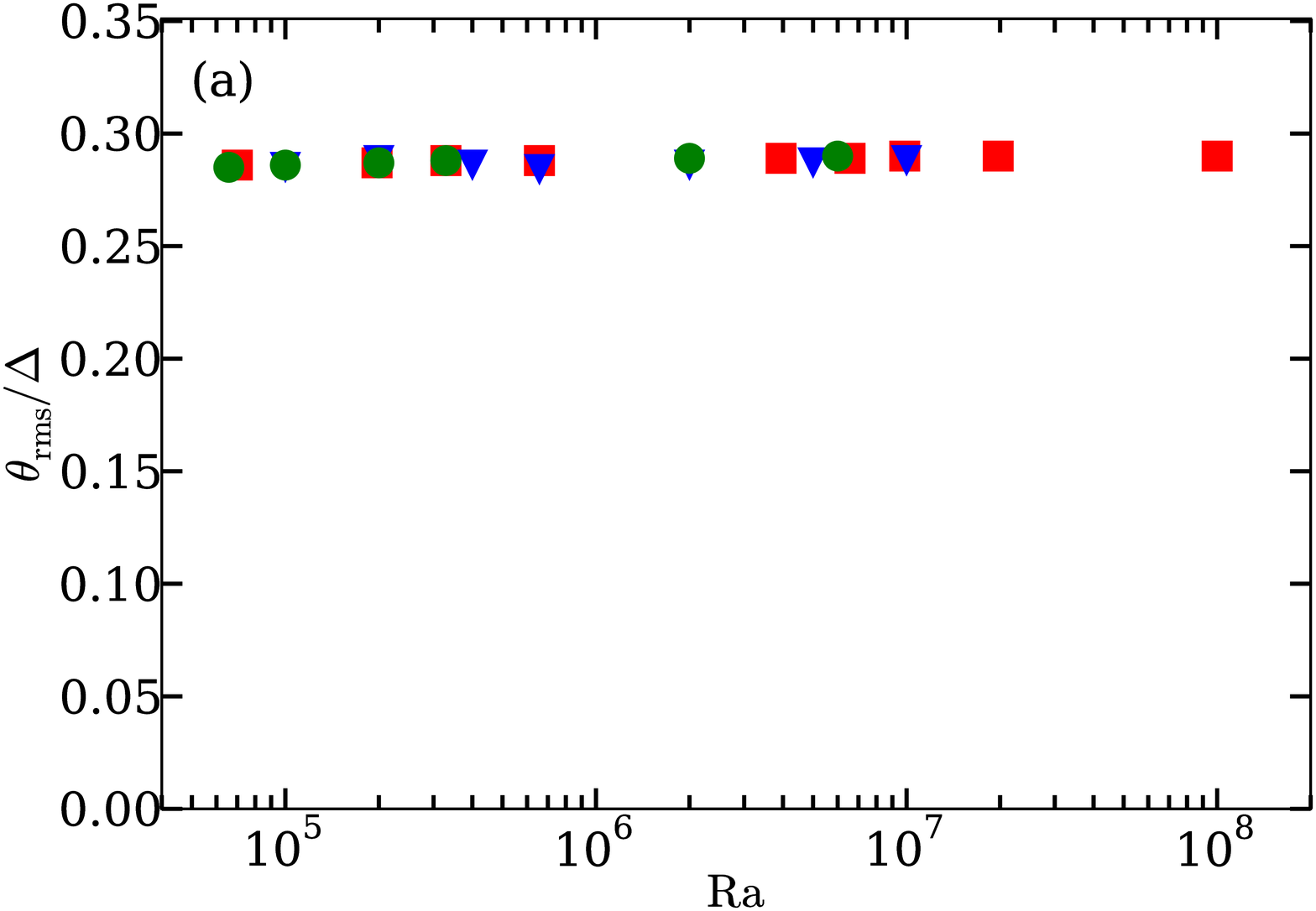}
\includegraphics[scale = 0.23]{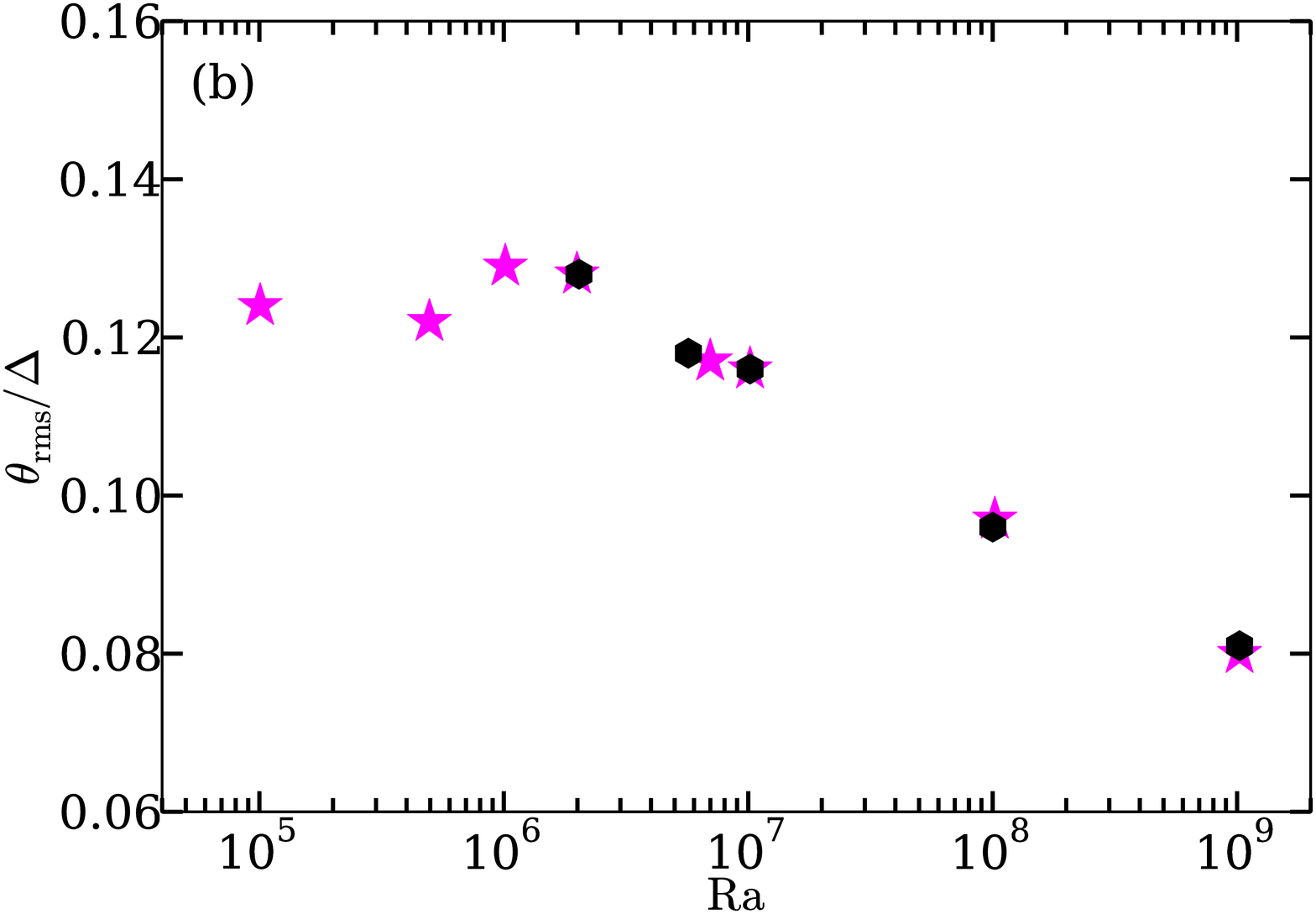}
\end{center}
\caption{(Color online) Plots of normalized root mean square thermal fluctuations ($\theta_{\mathrm{rms}}/\Delta$) vs.~Ra: (a) Upper panel: for our 3-d {\em free-slip runs} with $\mathrm{Pr}= \infty$ (red squares), $\mathrm{Pr}=10^2$ (blue down-pointing triangles), and $\mathrm{Pr}= 10^3$ (green circles); (b) Bottom panel:  for the 3-d no-slip simulations by Silano, \textit{et al.}~\cite{Silano:JFM2010} with $\mathrm{Pr}= 10^3$ (pink stars), and  $\mathrm{Pr}= 10^2$ (black hexagons).  $\theta_{\mathrm{rms}}/\Delta$ is an approximate constant for the free-slip boundary condition.  For the no-slip boundaries~\cite{Silano:JFM2010}, $\theta_{\mathrm{rms}}/\Delta$ is a constant for lower $\mathrm{Ra}$, and it is a weakly decreasing function of $\mathrm{Ra}$ for larger $\mathrm{Ra}$.}
\label{fig:th_ra}
\end{figure}

For our large Pr convection with free-slip boundary condition, we observe that $\theta_\mathrm{res} \ll \hat{\theta}(0,0,2)$.  Consequently, the rms fluctuation of $\theta$ is dominated by $\hat{\theta}(0,0,2)$ mode, thus yielding
\begin{equation}
\theta_\mathrm{rms}  \approx \sqrt{2 E_\theta} \approx \theta_L =  a_2 \Delta, \label{eq:theta_L_delta}
\end{equation}
which is independent of Ra and Pr, as depicted in Fig.~\ref{fig:th_ra}(a) for our free-slip runs. The constant $a_2 \approx (0.29 \pm 0.01)$ for free-slip runs with $\mathrm{Pr}=\infty, 10^2$ and $10^3$, which is reasonably close to the corresponding $\theta_\mathrm{rms}$ for intermediate Pr ($\mathrm{Pr}\sim 1$)~\cite{Verma:PRE2012}. For the no-slip condition however $\theta_\mathrm{rms}$ is smaller than that for the free-slip value, as shown by  Silano, \textit{et al.}~\cite{Silano:JFM2010} in their simulations (see Fig.~\ref{fig:th_ra}(b)).  In addition,   Silano, \textit{et al.}~\cite{Silano:JFM2010} also report that as a function of $\mathrm{Ra}$, $\theta_{\mathrm{rms}}$ remains constant for $\mathrm{Ra} \precsim 10^6$, but it  decreases very slowly for larger $\mathrm{Ra}$, with the power-law exponent smaller than 0.08~\cite{Silano:JFM2010}.

The aforementioned difference in the behaviour of $\theta_\mathrm{rms}$ for the two boundary conditions appears to be related to the thickness of boundary layers for these cases. Petschel, \textit{et al.}~\cite{Petschel:PRL2013} showed that the thermal boundary layer for the free-slip boundary condition is several times thinner than that for the no-slip boundary condition.  Hence the amplitudes of $\hat{\theta}(0,0,2n)$ modes for the no-slip boundary condition are expected to be smaller than those for the free-slip boundary condition, and $\theta_\mathrm{rms}$  could be comparable to the $\theta_\mathrm{res}$  for the no-slip condition. As a result,  $\theta_\mathrm{rms}$ for no-slip boundary condition is smaller than that for the free-slip boundary condition, as well as appear to show a weak decrease with $\mathrm{Ra}$ somewhat similar to $\theta_\mathrm{res}$ (see Fig.~\ref{fig:th_ra_2}).

\subsection{P\'{e}clet number scaling} \label{subsec:Pe}
\begin{figure}
\begin{center}
\includegraphics[scale = 0.23]{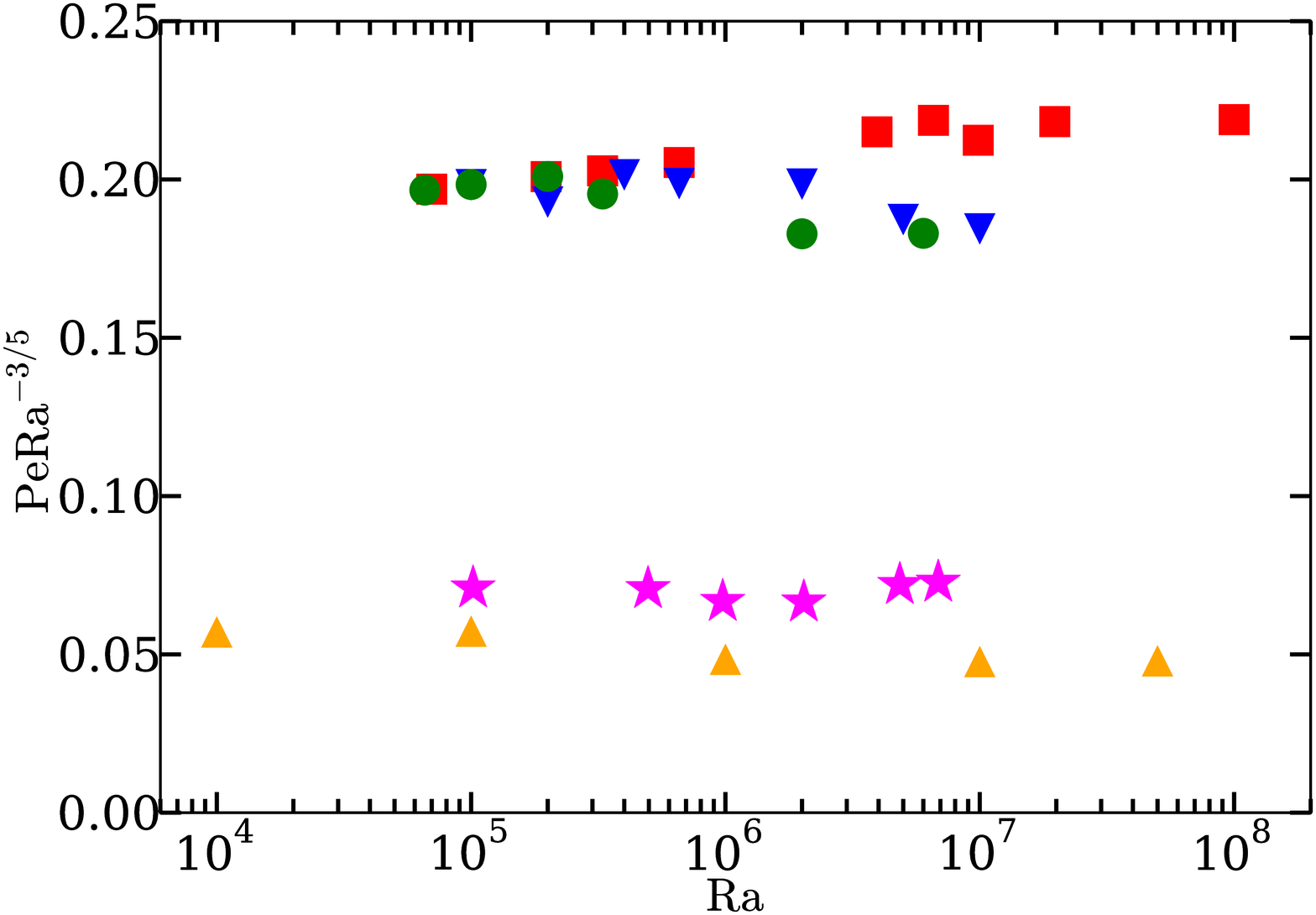}
\end{center}
\caption{(Color online) Plot of normalized P\'{e}clet number ($\mathrm{Pe}\mathrm{Ra}^{-3/5}$) vs.~Ra for the free-slip runs with $\mathrm{Pr}= \infty$ (red squares),  $\mathrm{Pr}= 10^2$ (blue down-pointing triangles),  and $\mathrm{Pr}= 10^3$ (green circles); and for the no-slip runs with $\mathrm{Pr}= 10^2$  (orange triangles) and  $\mathrm{Pr}= 10^3$ (pink stars; data from Silano, \textit{et al.}~\cite{Silano:JFM2010}). The prefactors for the no-slip runs are lower than those for the free-slip runs.}
\label{fig:pe_norm}
\end{figure}

Now we analyze the scaling of the P\'{e}clet number for large and infinite Prandtl number convection. Using our numerical data and Eq.~(\ref{eq:Pe}), we find that  
\begin{equation}
 \left( \sum_\mathbf k|\hat{\theta}(\mathbf k)|^2 \frac{k_\perp^2}{k^6} \frac{1}{ d^4 \Delta^2} \right)^{1/2} \approx \mathrm{Ra}^{-\zeta},
\end{equation} 
with $\zeta \approx 0.38$. Hence
\begin{equation}
\mathrm{Pe} = a_3 \mathrm{Ra}^{1-\zeta} =  a_3 \mathrm{Ra}^{0.62}. \label{eq:Pe_Ra}
\end{equation}
Similar relations are observed for $\mathrm{Pr}=10^2$ and $\mathrm{Pr}=10^3$, as shown in Fig.~\ref{fig:pe_norm} in which we plot $\mathrm{Pe} \mathrm{Ra}^{-3/5}$ vs. Ra. We find that $\mathrm{Pe} =(0.20 \pm 0.02)\mathrm{Ra}^{0.61 \pm 0.01}, (0.29 \pm 0.07)\mathrm{Ra}^{0.57 \pm 0.02}, (0.24 \pm 0.05)\mathrm{Ra}^{0.58 \pm 0.01}$ for $\mathrm{Pr}= \infty, 10^2$, and $10^3$ respectively.  For a no-slip simulation with $\mathrm{Pr}= 10^2$, $\mathrm{Pe} =(0.05 \pm 0.01)\mathrm{Ra}^{0.60 \pm 0.01}$. It is clear from Fig.~\ref{fig:pe_norm} that the prefactors for the no-slip runs (data from Silano, \textit{et al.}~\cite{Silano:JFM2010}) are smaller than those for the free-slip runs, which is due to the absence of wall friction for the free-slip boundary condition. Note that the wall friction slows down the flow further.   These results are in reasonable agreement with the earlier results of Silano, \textit{et al.}~\cite{Silano:JFM2010} (see Fig.~\ref{fig:pe_norm}), as well as with the GL scaling that $\mathrm{Pe} \approx 0.038 \mathrm{Ra}^{2/3}$  (the $\mathrm{Pr}=10^2$ no-slip data set belongs to the  I$_\infty^<$ regime). Another interesting aspect of the above scaling is its independence from Pr, unlike that for moderate Pr's ($\mathrm{Pr} \sim 1$) for which $\mathrm{Pe} \approx \sqrt{\mathrm{Ra}\mathrm{Pr}}$~\cite{Grossmann:JFM2000, Verma:PRE2012}.

Since the Reynolds number $\mathrm{Re}= \mathrm{Pe}/\mathrm{Pr}$, $\mathrm{Re}=0$ for $\mathrm{Pr}=\infty$, and $\mathrm{Re}$ is small for $\mathrm{Pr}\gg 1$. Hence the flow is viscous when the Prandtl number is large or infinite.  Note however that the Reynolds number tends to become larger than one (yet near one) for very large Ra (see Table~\ref{table:pr_inf}).

\subsection{Nusselt number scaling} \label{subsec:Nu_scaling}
Nusselt number  is defined as the ratio of total heat flux to the conductive heat flux, i.e., 
\begin{equation}
\mathrm{Nu} =  \frac{\kappa \Delta/d + \langle u_z T \rangle}{\kappa \Delta/d}  
= 1 + \langle \frac{u_z d}{\kappa} \frac{\theta_\mathrm{res}}{\Delta} \rangle
= 1 + \langle u'_z \theta'_{\mathrm{res}} \rangle,
\end{equation}
where $\theta'_{\mathrm{res}} = \theta_\mathrm{res}/\Delta$ is the normalized temperature fluctuation without $\hat{\theta}(0,0,2n)$ modes, and $u'_z = u_z d/\kappa$.  The absence of $(0,0,2n)$ Fourier modes in the above expression is due to the fact that $\hat{u}_z(0,0,2n)=0$ (see \textsection~\ref{sec:eqns}).

\begin{figure}
\begin{center}
\includegraphics[scale=0.23]{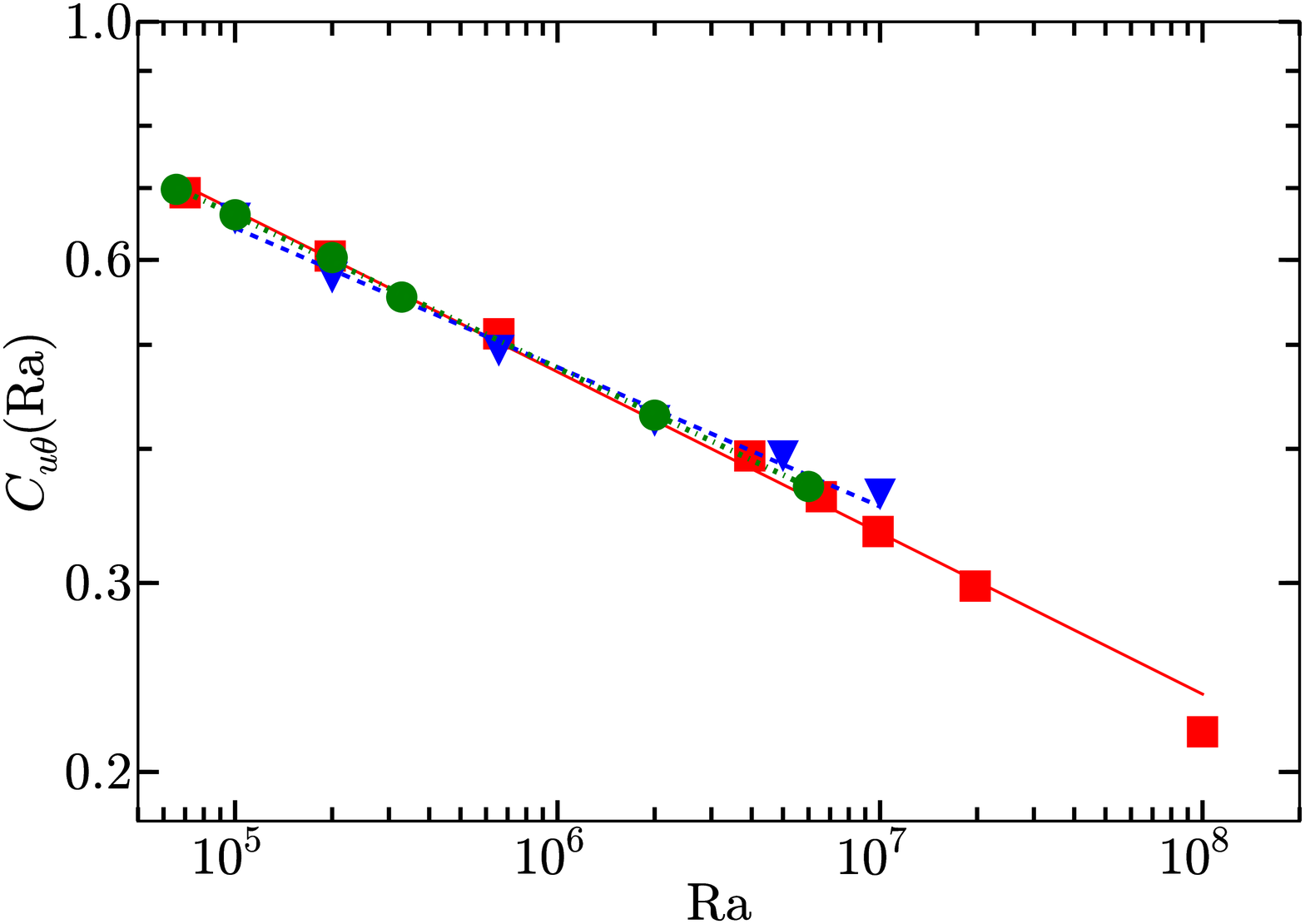}
\end{center}
\caption{(Color online) Plots of correlation function between the vertical velocity and temperature fields $C_{u\theta}(\mathrm{Ra})$ vs. Ra for the free-slip runs with $\mathrm{Pr}= \infty$ (red squares), $\mathrm{Pr}= 10^2$ (blue down-pointing triangles), and $\mathrm{Pr}= 10^3$ (green circles). $C_{u\theta} \approx a_4 \mathrm{Ra}^{-0.15}$.}
\label{fig:corr_uth}
\end{figure}

The above expression for Nu  can be rewritten as~\cite{Verma:PRE2012, Verma:IPR2013}
\begin{equation}
\mathrm{Nu} - 1 = \langle u_z' \theta'_{\mathrm{res}} \rangle = C_{u\theta}(\mathrm{Ra}) \langle u_z'^2 \rangle^{1/2}_V \langle\theta'^{2}_{\mathrm{res}}  \rangle_V^{1/2}, \label{eq:Nu_1}
\end{equation}
where the correlation function between the vertical velocity and temperature fields $C_{u\theta}(\mathrm{Ra})$ is
\begin{equation}
C_{u\theta}(\mathrm{Ra}) = \left \langle \frac{\langle u_z' \theta'_{\mathrm{res}} \rangle_V}{ \langle u_z'^2  \rangle_V^{1/2} \langle  \theta'^{2}_{\mathrm{res}} \rangle_V^{1/2} } \right \rangle_t. \label{eq:cuth}
\end{equation}
Here, $V$ and $t$ stand for the volume and temporal averages respectively. Our numerical data reveal that $C_{u\theta}(\mathrm{Ra}) = a_4 \mathrm{Ra}^{-0.15}$, as exhibited in Fig.~\ref{fig:corr_uth}. We observe that $C_{u\theta} = (3.8 \pm 0.6)\mathrm{Ra}^{-0.15 \pm 0.01}$ for $\mathrm{Pr}= \infty$,  $C_{u\theta} = (2.9 \pm 0.6)\mathrm{Ra}^{-0.13 \pm 0.02}$ for $\mathrm{Pr}=10^2$, and $C_{u\theta} = (3.4 \pm 0.2)\mathrm{Ra}^{-0.14 \pm 0.01}$ for $\mathrm{Pr}=10^3$.  

Eq.~(\ref{eq:Nu_1}) can be expressed as 
\begin{equation}
\mathrm{Nu}-1  \approx  \mathrm{Nu} \approx C_{u\theta}(\mathrm{Ra}) \langle  u_z'^2  \rangle^{1/2}_V \langle\theta'^2_{\mathrm{res}} \rangle_V^{1/2}.
\end{equation}
Using the scaling relations $C_{u\theta} = a_4\mathrm{Ra}^{-0.15}$, $\mathrm{Pe} = a_3 \mathrm{Ra}^{1-\zeta}$, and $\theta_{\mathrm{res}} = a_1 \mathrm{Ra}^{-\delta}$, we deduce that
\begin{equation}
\mathrm{Nu} = a_1 a_3 a_4  \mathrm{Ra}^{1 - \zeta - \delta - 0.15} = a_5 \mathrm{Ra}^{0.32},
\end{equation}
with $\zeta \approx 0.38$ and $\delta \approx 0.15$, and $a_5 = a_1 a_3 a_4$. Our arguments show that a subtle variations of Pe and $\theta'$ with respect to $\mathrm{Ra}$, and the correlation between the vertical velocity and temperature fields yield $\mathrm{Nu} \approx a_5 \mathrm{Ra}^{0.32}$. 

\begin{figure}
\begin{center}
\includegraphics[scale = 0.23]{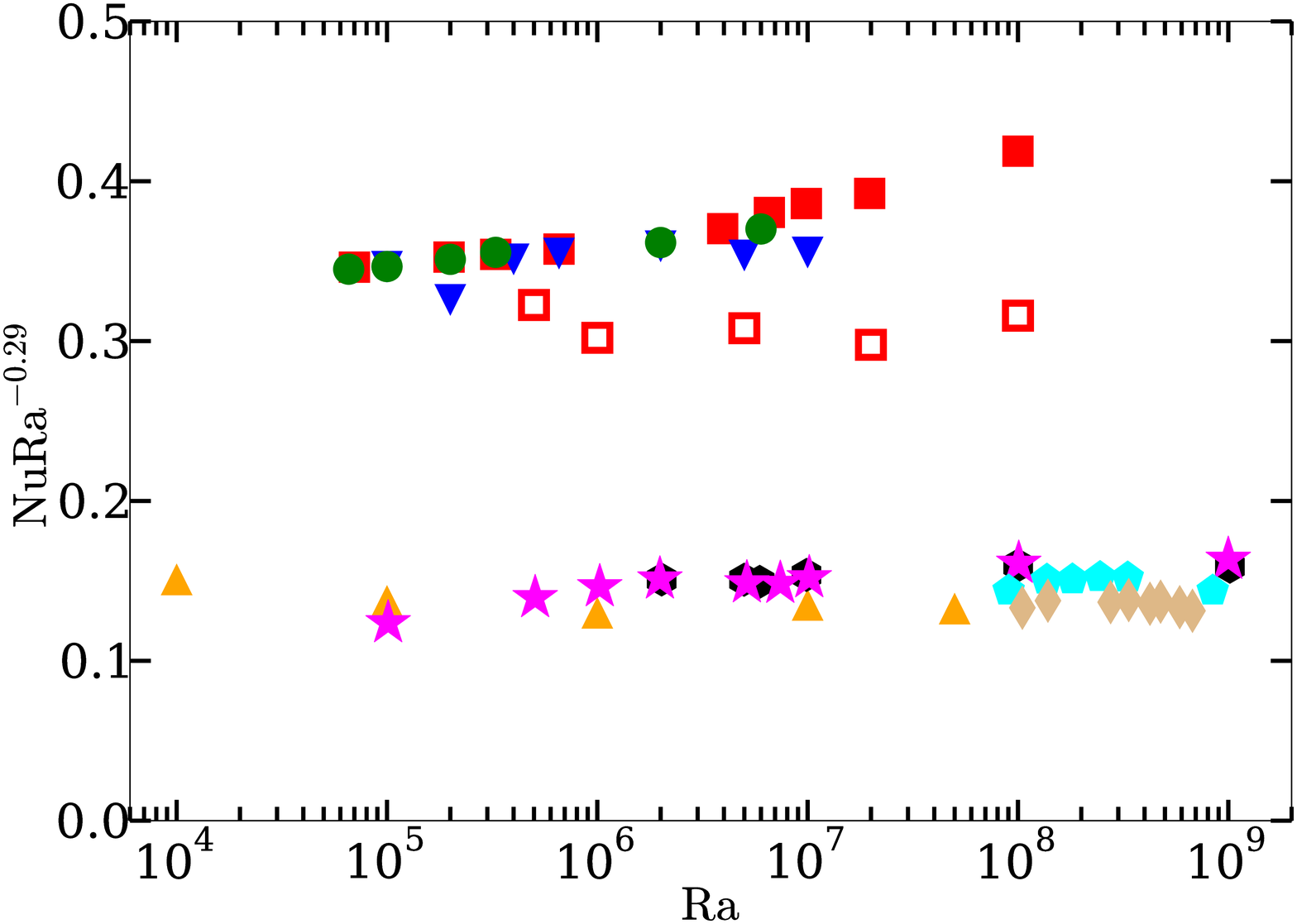}
\end{center}
\caption{(Color online) Normalized Nusselt number ($\mathrm{Nu}\mathrm{Ra}^{-0.29}$) as a function of Ra for free-slip runs with $\mathrm{Pr}= \infty$ (red squares),  $\mathrm{Pr}= 10^2$ (blue down-pointing triangles), $\mathrm{Pr}= 10^3$ (green circles), and for Pr = $\infty$ (2D) (open red squares, data taken from Hansen, \textit{et al.}~\cite{Hansen:PF1990}); and for no-slip runs with $\mathrm{Pr}= 10^2$ (2D) (orange triangles), $\mathrm{Pr}= 10^2$ (pink stars, data taken from Silano, \textit{et al.}~\cite{Silano:JFM2010}),  and $\mathrm{Pr}= 10^3$ (black hexagons, data taken from Silano, \textit{et al.}~\cite{Silano:JFM2010}).  Experimental data of Xia, \textit{et al.}~\cite{Xia:PRL2002} for $\mathrm{Pr}=205$ and $\mathrm{Pr}= 818$ are shown as cyan filled pentagons and burlywood diamonds respectively.}
\label{fig:nu_norm}
\end{figure} 

Scaling relations for $\mathrm{Pr}= \infty$, $10^2$, and $10^3$ (free-slip runs) computed using our numerical data are   $\mathrm{Nu} = (0.23 \pm 0.02)\mathrm{Ra}^{0.32 \pm 0.01}$, $\mathrm{Nu} =(0.32 \pm 0.07)\mathrm{Ra}^{0.30 \pm 0.01}$, and $\mathrm{Nu} =(0.29 \pm 0.01)\mathrm{Ra}^{0.31 \pm 0.01}$ respectively. In addition, for the no-slip run,  $\mathrm{Nu} =(0.14 \pm 0.03)\mathrm{Ra}^{0.29 \pm 0.01}$ for $\mathrm{Pr}= 10^2$. The prefactor for the free-slip runs is higher than that for the no-slip runs (data from Silano, \textit{et al.}~\cite{Silano:JFM2010} and Xia, \textit{et al.}~\cite{Xia:PRL2002}), which is reasonable since the heat transport is enhanced for the free-slip runs due to lower friction at the top and bottom plates (see Fig.~\ref{fig:nu_norm}). Also, for $\mathrm{Pr}= \infty$ with free-slip run, we observe that $a_1 \approx 0.58$, $a_3 \approx 0.20$, and $a_4 \approx 3.8$, and consequently $a_5 \approx 0.44$, which is in a reasonable agreement with the observed $a_5 \approx 0.23$.  Similar 
consistency is observed for $\mathrm{Pr}=10^2$ and $10^3$ as well.  Thus, our scaling results for $\theta$, Pe and Nu are consistent with each other.  

The aforementioned scaling results  are in good agreement with GL scaling~\cite{Grossmann:PRL2001}, according to which $\mathrm{Nu} = 0.17\mathrm{Ra}^{1/3}$ in I$_\infty^<$ regime. Our results are also consistent with the experimental results of Xia, \textit{et al.}~\cite{Xia:PRL2002} for Pr = 205 and  818, and the numerical results of Silano, \textit{et al.}~\cite{Silano:JFM2010} and Roberts~\cite{Roberts:GAFD1979} for large Pr simulations (see Fig.~\ref{fig:nu_norm}).   

\subsection{Scaling of dissipation rates} \label{subsec:dissipation_rates}
In Section~\ref{sec:eqns}, we derived relationships between the normalized dissipation rates and the Nusselt number. In this subsection we compute the normalized dissipation rates $C_{\epsilon_u}$, $C_{\epsilon_T,1}, C_{\epsilon_T,2}$ using numerical data and compare them with the exact results. 

\begin{figure}
\begin{center}
\includegraphics[scale=0.23]{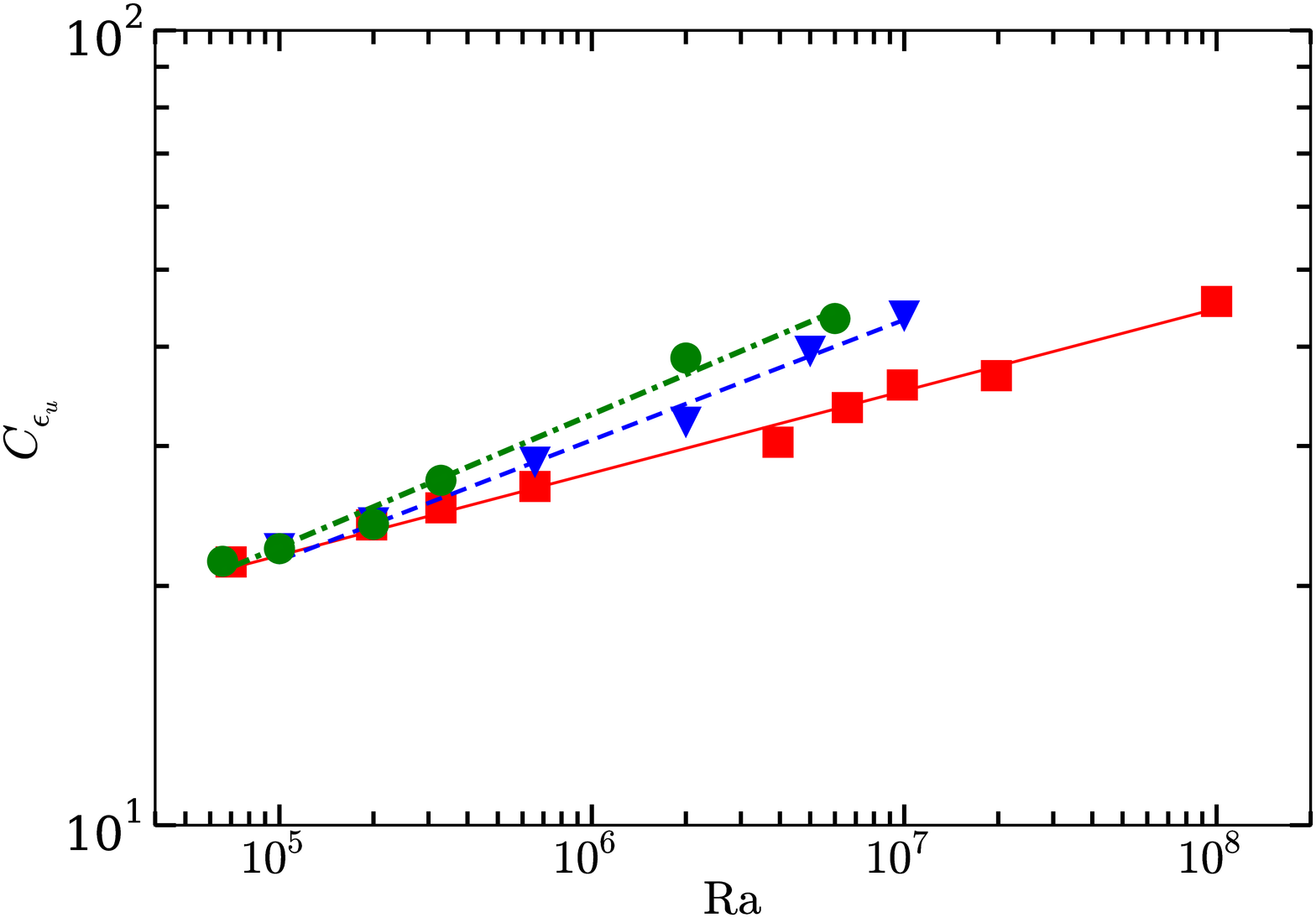}
\end{center}
\caption{(Color online) Plot of normalized viscous dissipation rate $C_{\epsilon_u}$ vs. Ra for the free-slip runs with $\mathrm{Pr}= \infty$ (red squares), $\mathrm{Pr}= 10^2$ (blue down-pointing triangles), and $\mathrm{Pr}= 10^3$ (green circles).  $C_{\epsilon_u}(\mathrm{Ra}) \sim  \mathrm{Ra}^{0.15}$, $\mathrm{Ra}^{0.17}$, and $\mathrm{Ra}^{0.10}$ for $\mathrm{Pr}= 10^2$, $10^3$, and $\infty$ respectively. }
\label{fig:corr_eps_u}
\end{figure}

From the exact relationship between the viscous dissipation rate and the Nusselt number (Eq.~(\ref{eq:C_eps_u})), we obtain
\begin{equation}
C_{\epsilon_u} = \frac{(\mathrm{Nu}-1) \mathrm{Ra}}{\mathrm{Pe}^2} \approx \frac{a_5}{a_3^2} \mathrm{Ra}^{\gamma+2\zeta-1} = \frac{a_5}{a_3^2} \mathrm{Ra}^{0.08}
\end{equation}
using $\gamma=0.32$ and $\zeta=0.38$.  For $\mathrm{Pr}=\infty$,  $a_5 \approx 0.23$ and $a_3 \approx 0.20$ provide the prefactor to be 5.8. Fig.~\ref{fig:corr_eps_u} shows the variation of $C_{\epsilon_u}$ with Ra for Pr = $10^2, 10^3$, and $\infty$. From our numerical data (Table~\ref{table:pr_inf}) we find that  $C_{\epsilon_u} = (3.8 \pm 1.2)\mathrm{Ra}^{0.15 \pm 0.02}$ for $\mathrm{Pr}= 10^2$, $C_{\epsilon_u} = (3.3 \pm 1.1)\mathrm{Ra}^{0.17 \pm 0.02}$ for $\mathrm{Pr}= 10^3$, and $C_{\epsilon_u} = (6.6 \pm 1.3)\mathrm{Ra}^{0.10 \pm 0.01}$ for $\mathrm{Pr}=\infty$.  These computed results are in good agreement with the aforementioned estimates using exact relationships, thus they validate our computations as well as show consistency with the other scaling relations.

\begin{figure}
\begin{center}
\includegraphics[scale=0.23]{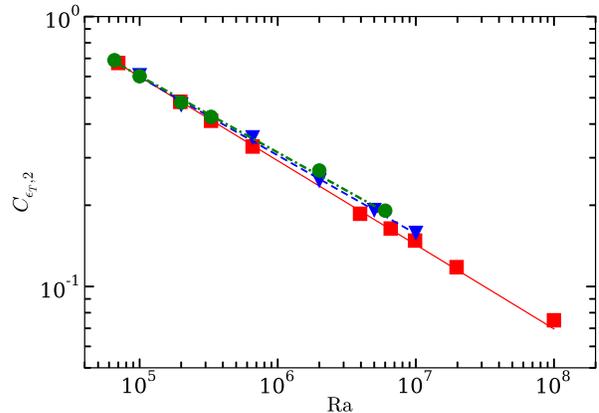}
\end{center}
\caption{(Color online) Plot of normalized thermal dissipation rate $C_{\epsilon_T,2}$ vs. Ra for the free-slip runs with $\mathrm{Pr}= \infty$ (red squares), $\mathrm{Pr}= 10^2$ (blue down-pointing triangles), and $\mathrm{Pr}= 10^3$ (green circles). $C_{\epsilon_T,2} \sim \mathrm{Ra}^{-0.29}$, $\mathrm{Ra}^{-0.28}$, and $\mathrm{Ra}^{-0.31}$ for $\mathrm{Pr}= 10^2$, $10^3$, and $\infty$ respectively.}
\label{fig:corr_eps_th_2}
\end{figure} 

According to Eq.~(\ref{eq:C_eps_th_1}), the  normalized thermal dissipation rate is defined as
\begin{equation}
C_{\epsilon_T,1} = \frac{\epsilon_T}{\kappa \Delta^2/d^2} = \mathrm{Nu} = a_5 \mathrm{Ra}^{\gamma}.
\end{equation}
Thus $C_{\epsilon_T,1}$ has a same scaling as the Nusselt number (see Table~\ref{table:pr_inf}). The scaling of the other normalized thermal dissipation rate is more complex.  Eq.~(\ref{eq:C_eps_th_2}) yields
\begin{equation}
C_{\epsilon_T,2} = \frac{\epsilon_T}{U_L \theta_L^2/d} = \frac{\mathrm{Nu}}{\mathrm{Pe}} \left( \frac{\Delta}{\theta_L} \right)^2.
\end{equation}
Using the scaling of Nu, Pe and $\theta_L$, we deduce
\begin{equation}
C_{\epsilon_T,2} = \frac{a_5}{a_3 a_2^2} \mathrm{Ra}^{\gamma + \zeta - 1} = \frac{a_5}{a_3 a_2^2} \mathrm{Ra}^{-0.30},
\end{equation}
with $\gamma = 0.32$ and $\zeta = 0.38$.  Using the constants $a_i$'s, the prefactor is approximately 14 for $\mathrm{Pr}=\infty$.  Similar exponents and prefactors are observed for other large Pr simulations. In Fig.~\ref{fig:corr_eps_th_2} we plot $C_{\epsilon_T,2}$ vs. Ra, which exhibits $C_{\epsilon_T,2} = (17 \pm 5.1)\mathrm{Ra}^{-0.29 \pm 0.02}$ for $\mathrm{Pr}= 10^2$, $C_{\epsilon_T,2} = (16 \pm 4.1)\mathrm{Ra}^{-0.28 \pm 0.02}$ for $\mathrm{Pr}= 10^3$, and $C_{\epsilon_T,2} = (22 \pm 2.2)\mathrm{Ra}^{-0.31 \pm 0.01}$ for $\mathrm{Pr}=\infty$. These numerical results are in good agreement with the theoretical estimates computed earlier. For Pr = 0.7, Emran and Schumacher~\cite{Emran:JFM2008, Emran:EPJE2012} also observed nearly similar scaling for thermal dissipation rate. They estimated thermal dissipation rates separately in the bulk and boundary layers, as well as in the plume-dominated regions and in the turbulent background.

The aforementioned scaling of the dissipation rates and their consistency with other global quantities like Nu and Pe indicate consistency of our arguments.  These results are summarized in Table~\ref{table:summary_inf}  for a free-slip simulation with $\mathrm{Pr}=\infty$, and in Table~\ref{table:summary_noslip}  for a no-slip simulation with $\mathrm{Pr}= 10^2$. The exponents and the prefactors are nearly the same for all $\mathrm{Pr}\ge 10^2$. 

After our discussion on the global quantities, we turn to the computations of energy and entropy spectra, as well as their fluxes.

\section{Energy and entropy spectra} \label{sec:spectrum}
Energy and entropy contained in a wavenumber shell of radius $k$ are called the energy spectrum $E_u(k)$ and entropy spectrum $E_\theta(k)$ respectively, i.e.,
\begin{eqnarray}
E_u(k) & = & \sum_{k \leq |{\bf k'}| < k+1} \frac{|\hat{\mathbf u}(\mathbf k')|^2}{2},  \\
E_{\theta}(k) & = & \sum_{k \leq |{\bf k'}| < k+1} \frac{|\hat{\theta}(\mathbf{k'})|^2}{2}.
\end{eqnarray}
Nonlinear interactions lead to a transfer of energy and entropy from smaller wavenumber modes to larger wavenumber modes.  These transfers are quantified using energy flux $\Pi_u(k_0)$ and entropy flux $\Pi_\theta(k_0)$, which are the fluxes coming out of a wavenumber sphere of radius $k_0$~\cite{Verma:PR2004,Mishra:PRE2010} 
\begin{equation}
\Pi_u(k_0) = \sum_{k \geq k_0} \sum_{p < k_0} \delta_{\mathbf k, \mathbf p+\mathbf q} \Im([\mathbf k \cdot \hat{\mathbf u}(\mathbf q)][\hat{\mathbf u}^*(\mathbf k) \cdot \hat{\mathbf u}(\mathbf p)]),
\end{equation}
\begin{equation}
\Pi_{\theta}(k_0) = \sum_{k \geq k_0} \sum_{p < k_0} \delta_{\mathbf k, \mathbf p+\mathbf q} \Im([\mathbf k \cdot \hat{\mathbf u}(\mathbf q)][\hat{\theta}^*(\mathbf k) \cdot \hat{\theta} (\mathbf p)]),
\end{equation}
where $ \Im$ is the imaginary part of the argument, and ${\mathbf k,\mathbf p,\mathbf q}$ are the wavenumbers of a triad with ${\mathbf k=\mathbf p+ \mathbf q}$.  

For $\mathrm{Pr}= \infty$, the momentum equation (Eq.~(\ref{eq:u_infty})) yields
\begin{equation}
\alpha g (\theta_\mathrm{res})_l \approx \frac{\nu u_l}{l^2}.
\end{equation}
We take $(\theta_{\mathrm{res}})_l = \mathrm{Ra}^{-\delta} \theta_l$ with $\delta$ defined in Eq.~(\ref{eq:theta_res_delta}). We also assume a constant entropy flux, which yields
\begin{equation}
\epsilon_\theta = \frac{\theta_L^2 U_L}{d}  C_{\epsilon_\theta,2} = \frac{\theta_l^2 u_l}{l} C_{\epsilon_\theta,2}.
\end{equation}
We use the expressions of $\theta_L$ and $U_L$ derived earlier  (Eqs.~(\ref{eq:theta_L_delta}) and (\ref{eq:Pe_Ra})).  After substitutions of these expressions in the above equations we obtain
\begin{eqnarray}
u_l & = & (a_2^2  a_3)^{\frac{1}{3}} \frac{\kappa}{d} \mathrm{Ra}^{\frac{1}{3}(3-2\delta-\zeta)} \left( \frac{l}{d} \right)^{\frac{5}{3}}, \\
\theta_l & = & (a_2^2  a_3)^{\frac{1}{3}} \Delta \mathrm{Ra}^{\frac{1}{3}(\delta-\zeta)} \left( \frac{l}{d} \right)^{-\frac{1}{3}},
\end{eqnarray}
and therefore the energy and entropy spectra are
\begin{eqnarray}
E_u(k) & = & (a_2^2  a_3)^{\frac{2}{3}} d \left(\frac{\kappa}{d}\right)^2 \mathrm{Ra}^{\frac{2}{3}(3-2\delta-\zeta)}(kd)^{-\frac{13}{3}}, \\
E_\theta(k) & = & (a_2^2  a_3)^{\frac{2}{3}} d \Delta^2 \mathrm{Ra}^{\frac{2}{3}(\delta-\zeta)} (kd)^{-\frac{1}{3}}.
\end{eqnarray}
We define normalized spectra $\tilde{E}_u(k)$ and  $\tilde{E}_\theta(k)$ as 
\begin{eqnarray}
\tilde{E}_u(k)  & = & \frac{E_u(k)}{d \left(\frac{\kappa}{d}\right)^2 \mathrm{Ra}^{\frac{2}{3}(3-2\delta-\zeta)}} = (a_2^2  a_3)^{\frac{2}{3}} (kd)^{-\frac{13}{3}}, \\
\tilde{E}_\theta(k) & = & \frac{E_\theta(k)}{d \Delta^2 \mathrm{Ra}^{\frac{2}{3}(\delta-\zeta)}} = (a_2^2  a_3)^{\frac{2}{3}} (kd)^{-\frac{1}{3}}.
\end{eqnarray}
For $\mathrm{Pr}=\infty$, the prefactor $(a_2^2  a_3)^{2/3}$ computed using $a_i$'s discussed earlier is approximately $6.3 \times 10^{-2}$.

\begin{figure}
\begin{center}
\includegraphics[scale = 0.23]{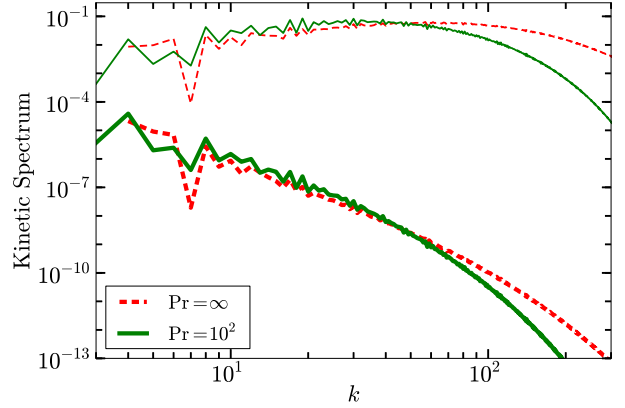}
\end{center}
\caption{(Color online) Kinetic energy spectrum $\tilde{E}_u(k)$ for $\mathrm{Pr}= \infty, \mathrm{Ra}= 10^8$  (thick red dashed curve), and for $\mathrm{Pr}= 10^2, \mathrm{Ra}= 10^7$ (thick green curve).  The respective normalized kinetic spectra $\tilde{E}_u(k)k^{13/3}$ (thin curves)  are approximate constants in the powerlaw range, thus $E_u(k) \sim k^{-13/3}$.}
\label{fig:Ek_freeslip}
\end{figure}

\begin{figure}
\begin{center}
\includegraphics[scale = 0.23]{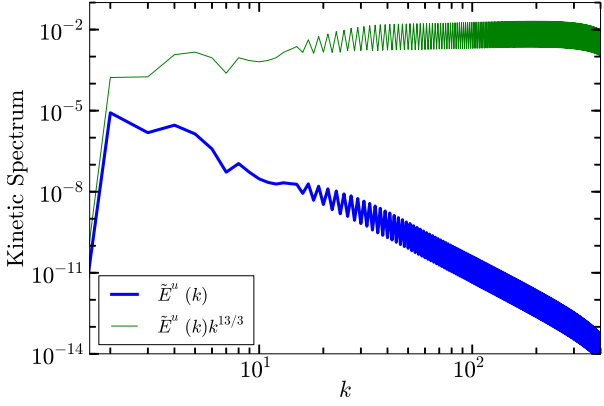}
\end{center}
\caption{(Color online) For no-slip box with $\mathrm{Pr}= 10^2$, and $\mathrm{Ra}= 10^7$: the kinetic energy spectrum $\tilde{E}_u(k)$ (thick blue curve), and the compensated kinetic energy spectrum $\tilde{E}_u(k)k^{13/3}$ (thin green curve).  The scaling is similar to that for the free-slip runs, except for a smaller prefactor. } 
\label{fig:ks_p100_no}
\end{figure}

In Fig.~\ref{fig:Ek_freeslip} we plot kinetic spectrum $\tilde{E}_u(k)$ and normalized kinetic spectrum $\tilde{E}_u(k)k^{13/3}$ for $(\mathrm{Pr}= \infty, \mathrm{Ra}= 10^8)$ and $(\mathrm{Pr}= 10^2, \mathrm{Ra}= 10^7)$. Normalized kinetic spectrum appears nearly constant for more than one decade of wavenumber with $\tilde{E}_u(k) \approx (0.06 \pm 0.02) k^{-13/3}$.   This result is in a very good agreement with the predictions based on scaling arguments.  Note that the $E_u(k)$ follows neither  the Bolgiano-Obukhov  nor the Kolmogorov-Obukhov scaling~\cite{Lvov:PD1992, Mishra:PRE2010}  since the velocity field is viscous. The above scaling law for the kinetic energy spectrum also holds for $\mathrm{Pr}=10^3$ for free-slip boundary condition, and for $\mathrm{Pr}=10^2$ for no-slip boundary condition, as exhibited in Fig.~\ref{fig:ks_p100_no}. For better statistics, we average approximately 35 frames for free-slip data;  the no-slip data however is not averaged. The prefactors  for the free-slip runs are larger than those for the no-slip run due to lower frictional forces for the free-slip convection. We however remark that  the data points for the no-slip boundary condition are not uniformly distributed, which necessitates interpolation of the data points to a uniform mesh. It is possible that the sawtooth-like spectrum at high wavenumbers could be due to the interpolation process.  Also, the sharp viscous boundary layer for the no-slip box could  produce fluctuations in the spectrum.

\begin{figure}
\begin{center}
\includegraphics[scale = 0.23]{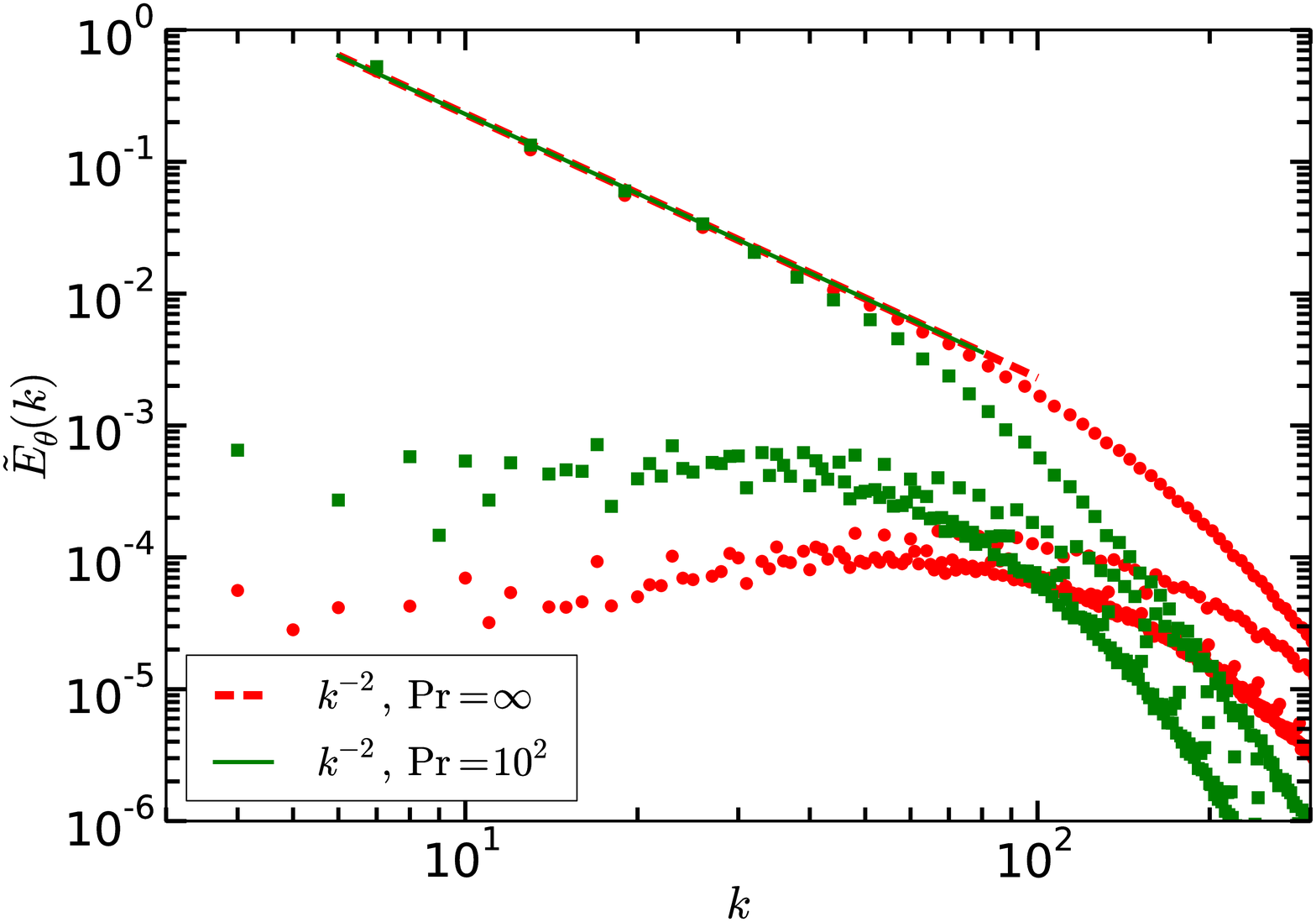}
\end{center}
\caption{(Color online) Plot of entropy spectrum $\tilde{E}_\theta(k)$ vs. $k$ for $\mathrm{Pr}= \infty, \mathrm{Ra}= 10^8$  (red filled circles) and for $\mathrm{Pr}= 10^2, \mathrm{Ra}= 10^7$ (green filled squares). The upper branches of the spectra represent  $\hat{\theta}(0,0,2n)$ modes that exhibit $k^{-2}$ scaling (red dashed and green solid curves). The lower branch is somewhat flat. }
\label{fig:Eth_freeslip}
\end{figure}

\begin{figure}
\begin{center}
\includegraphics[scale = 0.23]{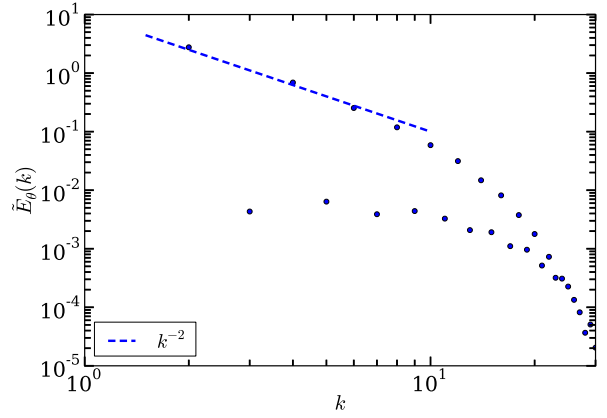}
\end{center}
\caption{(Color online) Plot of entropy spectrum $\tilde{E}_\theta(k)$ for $\mathrm{Pr}= 10^2, \mathrm{Ra}= 10^7$ with no-slip boundary condition. The spectrum exhibits similar scaling as large-Pr free-slip runs.}
\label{fig:ts_p100_no}
\end{figure}

The entropy spectrum, however, is more complex.   The entropy spectrum $\tilde{E}_\theta(k)$  exhibited in Fig.~\ref{fig:Eth_freeslip} for $(\mathrm{Pr}= \infty, \mathrm{Ra}= 10^8)$ and $(\mathrm{Pr}= 10^2, \mathrm{Ra}= 10^7)$ exhibits dual branches. The upper branch, which corresponds to the $\hat{\theta}(0,0,2n)$ Fourier modes, follows  $k^{-2}$ energy spectrum since $\hat{\theta}(0,0,2n) \approx -1/(2n\pi)$ (see \textsection~\ref{sec:eqns} and~\cite{Mishra:PRE2010}). The lower branch is the energy spectrum of the Fourier modes other than $\hat{\theta}(0,0,2n)$, and it follows nearly a flat spectrum. The nature of the entropy spectrum is very different from our phenomenological predictions that $E_{\theta}(k) \propto k^{-1/3}$.  This discrepancy is due to the boundary condition (the conducting plates) which yields significant $\hat{\theta}(0,0,2n)$ branch (see \textsection~\ref{sec:eqns} and Mishra and Verma~\cite{Mishra:PRE2010}). Similar behavior is observed for $\mathrm{Pr}=10^3$ with the free-slip 
boundary condition, and for $\mathrm{Pr}=10^2$ with no-slip boundary condition, as exhibited in Fig.~\ref{fig:ts_p100_no}.

\begin{figure}
\begin{center}
\includegraphics[scale = 0.23]{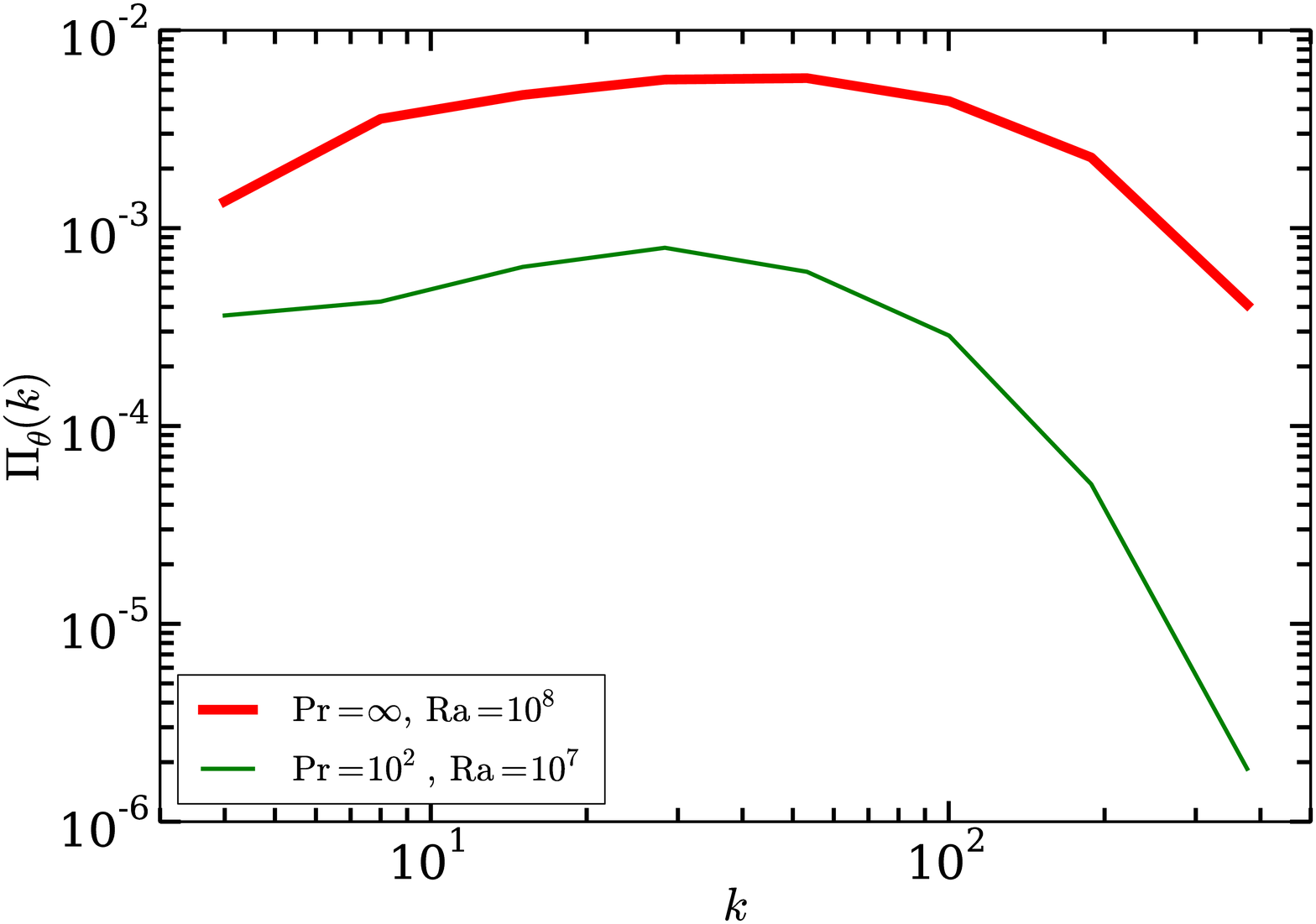}
\end{center}
\caption{(Color online) Plot of entropy flux $\Pi_{\theta}(k)$ vs. $k$ for $\mathrm{Pr}=\infty, \mathrm{Ra}= 10^8$ (thick red curve), and for  $\mathrm{Pr}= 10^2, \mathrm{Ra}= 10^7$ (thin green curve). The fluxes are approximate constants in the powerlaw range.}
\label{fig:flux}
\end{figure}

We also compute the entropy flux $\Pi_\theta(k_0)$ using the numerical data. In  Fig.~\ref{fig:flux} we plot $\Pi_{\theta}(k)$ for $\mathrm{Pr}= \infty$ and $\mathrm{Ra}= 10^8$, and for $\mathrm{Pr}= 10^2$ and $\mathrm{Ra}= 10^7$.  The plots indicate a nearly constant entropy flux in the powerlaw regime. The constancy of $\Pi_{\theta}(k)$ is due to the dominance of nonlinear term in the temperature equation. The kinetic energy flux $\Pi_{u}(k)$ however is zero for $\mathrm{Pr}=\infty$ due to the absence of  the nonlinearity  in the velocity equation.  For large Pr runs, $\Pi_{u}(k)$ is very small due to weak nonlinearity.

In Table~\ref{table:summary_inf} we summarize the scaling results for $\mathrm{Pr}=\infty$ simulation under free-slip boundary condition, and Table~\ref{table:summary_noslip}  for $\mathrm{Pr}=10^2$ simulation under no-slip boundary condition.  Here we list the theoretically-estimated and numerically-computed values. The two sets of values are in good agreement with each other.  The scaling for large Pr ($\mathrm{Pr}\ge 10^2$) RBC  is very similar to that for $\mathrm{Pr}=\infty$ RBC.  Our data for no-slip boundary condition are somewhat limited at present.

\begin{table*}
\begin{center}
\caption{Summary of the scaling functions for various quantities for free-slip runs with $\mathrm{Pr}= \infty$. The estimated values using analytical arguments agree quite well with the numerically computed ones.  The formulas for the scaling functions using $a_i$'s are listed in the second column.}
\begin{tabular*}{\textwidth}{p{4cm} p{4cm} p{5cm} p{5cm}}
\hline \hline \\[1 pt]
Quantity & Formula & Estimated & Computed  \\[2 mm]
\hline \\[1 pt]
$\theta_{\mathrm{res}}/\Delta$ & $a_1 \mathrm{Ra}^{-\delta}$ & $-$ & $(0.59 \pm 0.08)\mathrm{Ra}^{-0.15 \pm 0.01}$ \\
$\theta_{\mathrm{rms}}/\Delta$ & $a_2$ & $-$ & $(0.29 \pm 0.01) $ \\
Pe & $a_3 \mathrm{Ra}^{1-\zeta}$ & $0.16\mathrm{Ra}^{0.62}$ & $(0.20 \pm 0.02)\mathrm{Ra}^{0.61 \pm 0.01}$ \\
$C_{u\theta}$ & $a_4 \mathrm{Ra}^{-0.15}$ & $-$ & $(3.8 \pm 0.6)\mathrm{Ra}^{-0.15 \pm 0.01}$ \\
Nu & $a_5 \mathrm{Ra}^{\gamma}$ & $0.43\mathrm{Ra}^{0.32}$ & $(0.23 \pm 0.02)\mathrm{Ra}^{0.32 \pm 0.01}$ \\
$C_{\epsilon_u}$ & $(a_5/a_3^2) \mathrm{Ra}^{\gamma+2\zeta-1}$ & $5.8\mathrm{Ra}^{0.08}$ & $(6.6 \pm 1.3)\mathrm{Ra}^{0.10 \pm 0.01}$ \\
$C_{\epsilon_T,2}$ & $a_5/(a_3 a_2^2) \mathrm{Ra}^{\gamma+\zeta-1}$ & $14\mathrm{Ra}^{-0.30}$ & $(22 \pm 2.2)\mathrm{Ra}^{-0.31 \pm 0.01}$ \\
$\tilde{E}_u(k)$ & $(a_2^2 a_3)^{2/3} k^{-13/3}$ & $0.063k^{-13/3}$ & $(0.06 \pm 0.02)k^{-13/3}$ \\
$\tilde{E}_\theta(k)$ & $(a_2^2 a_3)^{2/3} k^{-1/3}$ & $0.063k^{-1/3}$ & Dual branches \\
\hline \hline
\end{tabular*}
\label{table:summary_inf}
\end{center}
\end{table*}

\begin{table}
\begin{center}
\caption{Summary of the scaling functions for various quantities for no-slip runs with $\mathrm{Pr}=10^2$. The computed $\theta_{\mathrm{rms}}$ is taken from Silano, \textit{et al.}~\cite{Silano:JFM2010}. The estimated Pe and Nu scaling are taken from Grossmann and Lohse's~\cite{Grossmann:PRL2001} theoretical work (the I$^<_\infty$ regime).}
\begin{tabular}{p{2cm} p{2.5cm} p{3.5cm}}
\hline \hline \\[1 pt]
Quantity & Estimated & Computed \\[2 mm]
\hline \\[1 pt]
$\theta_{\mathrm{rms}}$ & $-$ & $(0.12 \pm 0.02) \Delta$ \\
Pe & $0.038\mathrm{Ra}^{2/3}$ & $(0.05 \pm 0.01)\mathrm{Ra}^{0.60 \pm 0.01}$ \\
Nu & $0.17\mathrm{Ra}^{1/3}$ & $(0.14 \pm 0.03)\mathrm{Ra}^{0.29 \pm 0.01}$ \\
$\tilde{E}_u(k)$ & $0.008 k^{-13/3}$ & $(0.006 \pm 0.004) k^{-13/3}$ \\
$\tilde{E}_\theta(k)$ & $0.008 k^{-1/3}$ & Dual branches \\
\hline \hline
\end{tabular}
\label{table:summary_noslip}
\end{center}
\end{table}

\section{Conclusions and discussions} \label{sec:conclusion}
In this paper we derive scaling properties of the large-scale quantities (e.g., P\'{e}clet and Nusselt numbers), as well as that of energy and entropy spectra for very large and infinite Prandtl number convection.  The equation for the velocity field is linear for $\mathrm{Pr}=\infty$ limit that helps us derive relationships between various quantities.  These scaling relations are verified using numerical simulations  for infinite and large  Prandtl numbers ($\mathrm{Pr}\ge 10^2$).   We observe that the scaling of flows with large Prandtl number ($\mathrm{Pr}\ge 10^2$) is same as that for the infinite Prandtl number, thus making them $\mathrm{Pr}$-independent.  Our results are consistent with the earlier theoretical predictions of Grossmann and Lohse~\cite{Grossmann:PRL2001}, experimental results of Xia, \textit{et al.}~\cite{Xia:PRL2002}, and the numerical results of Silano, \textit{et al.}~\cite{Silano:JFM2010}.  Note however that the analytical work of Grossmann and Lohse~\cite{Grossmann:PRL2001} is based 
on Shraiman and Siggia's~\cite{Shraiman:PRA1990} exact relations and modelling of the dissipation rates at the bulk and boundary layers.  While our theoretical work is based on the dimensional and scaling analysis of the dynamical equation of the velocity and temperature fields, as well as several inputs from the numerical simulations.

A summary of our results is as follows.
\begin{enumerate}
\item The temperature field is dominated by the Fourier modes $\hat{\theta}(0,0,2n)$, which are approximately $-1/(2n\pi)$ for small $n$ in accordance with the predictions by Mishra and Verma~\cite{Mishra:PRE2010}.  The modes other than $\hat{\theta}(0,0,2n)$ are termed as ``residual modes" whose rms fluctuations scale as $\mathrm{Ra}^{-\delta}$ with $\delta \approx 0.15$.  Due to the dominance of $\hat{\theta}(0,0,2n)$ modes, the large-scale temperature fluctuations follow $\theta_\mathrm{rms} \sim \Delta$ for the free-slip boundary condition, where $\Delta$ is the temperature difference between the hot and cold plates.   However the numerical results of Silano, \textit{et al.}~\cite{Silano:JFM2010} for the no-slip boundary condition exhibit the above behaviour  (with a smaller prefactor) for lower $\mathrm{Ra}$, but $\theta_\mathrm{rms}$ appear to decrease slowly with $\mathrm{Ra}$ for larger $\mathrm{Ra}$. We show that the residual modes play a very important role in the scaling of Nusselt 
number, energy spectrum, etc.

\item The P\'{e}clet number, which is proportional to the large-scale velocity, scales as $\mathrm{Ra}^{1-\zeta}$ with $\zeta \approx 0.38$.    Note that the Reynolds number in the large Pr limit is small, i.e., $\mathrm{Re} \lessapprox 1$.  These results are consistent with the theoretical predictions of GL~\cite{Grossmann:PRL2001} and the numerical results of Silano, \textit{et al.}~\cite{Silano:JFM2010}. 

\item The Nusselt number scales as $\mathrm{Ra}^{\gamma}$ with the exponent lying in the range from 0.30 to 0.32, which is consistent with the results of Grossmann and Lohse~\cite{Grossmann:PRL2001}, Silano, \textit{et al.}~\cite{Silano:JFM2010}, Roberts~\cite{Roberts:GAFD1979}, Xia, \textit{et al.}~\cite{Xia:PRL2002}, and Constantin and Doering~\cite{Constantin:JSP1999}. This scaling arises due to a complex interplay between the residual modes, P\'{e}clet number, and the velocity-temperature correlation. 

\item The normalized viscous and thermal dissipation rates are functions of Ra.  We observe that
\begin{eqnarray}
C_{\epsilon_u} & = & \frac{\epsilon_u}{\nu U_L^2/d^2} \approx \mathrm{Ra}^{b_1}, \\
C_{\epsilon_T,2} & = & \frac{\epsilon_T}{U_L \theta_L^2/d} \approx \mathrm{Ra}^{-b_2},
\end{eqnarray}
with $b_1 \approx 0.10$ and $b_2 \approx 0.31$.  These relations are consistent with the Nu scaling derived using the exact relations of Shraiman and Siggia~\cite{Shraiman:PRA1990}.  Here we derive an explicit Ra-dependent normalized dissipation rates for large Pr for the first time.

\item Using analytical arguments, we derive that the energy spectrum $E_u(k) \sim k^{-13/3}$.  Our simulations verify this power law in the powerlaw range for both the free-slip and no-slip boundary conditions.  

\item We predict that the entropy spectrum $E_\theta(k) \sim k^{-1/3}$.  Unfortunately this power law is not observed in the numerical simulations.  Instead, we find dual entropy spectra consisting of  an upper branch with $k^{-2}$ spectrum corresponding to $\hat{\theta}(0,0,2n) \approx -1/(2 n \pi)$, and a nearly flat lower branch.  The dual branching is due to the presence of boundary layers~\cite{Mishra:PRE2010}.

\item Our numerical simulations  show that the free-slip and no-slip boundary conditions provide similar scaling relations for the global quantities, as well as for the  energy and entropy spectra.  However, the prefactors of the P\'{e}clet and Nusselt numbers, and that of energy spectrum are smaller for the no-slip condition than those for the free-slip boundary condition. This discrepancy is due to a smaller frictional force experienced by the flow for the free-slip boundary condition. The similarities of the scaling functions between the free-slip and no-slip convection are due to the dominance of the large-scale flows, which have similar structures for both the free-slip and no-slip boundary conditions.  Note that viscous boundary layers pervade the whole box for $\mathrm{Pr}=\infty$, hence they determine the properties of the bulk flow.
\end{enumerate}

In summary, we derived scaling relations for large-scale quantities, and energy and entropy spectra for large and infinite Prandtl number convection.  The scaling properties are independent of the Prandtl number in this regime.  Our analytical and numerical results are consistent with  earlier results of Grossmann and Lohse~\cite{Grossmann:PRL2001}, Silano, \textit{et al.}~\cite{Silano:JFM2010}, and Xia, \textit{et al.}~\cite{Xia:PRL2002}.  
 
\section*{Acknowledgement} 

Our numerical simulations were performed at {\em hpc} and {\em Chaos} clusters of IIT Kanpur. This work was supported through the Swarnajayanti fellowship to MKV from Department of Science and Technology, India. We thank K. Sandeep Reddy, Mani Chandra, and Biplab Dutta for helpful tips on Matplotlib and NEK5000 softwares.

\end{document}